\documentclass[5p,final]{elsarticle} 
\journal{Journal of Sensors and Actuators A: Physical}
\usepackage{lineno,hyperref}
\modulolinenumbers[5]
\usepackage[acronym]{glossaries} 
\biboptions{sort&compress}
\usepackage[normalem]{ulem}
\usepackage{booktabs}
\usepackage{caption}
\usepackage{multirow}
\usepackage{floatrow}
\usepackage{float,longtable}
\usepackage[table]{xcolor}
\usepackage{subfig}
\usepackage{array}
\usepackage{gensymb}
\usepackage{amsmath}
\usepackage{dblfloatfix}
\usepackage{threeparttable}
\usepackage{xurl}
\usepackage{graphicx}
\newcolumntype{x}[1]{>{\centering\arraybackslash\hspace{0pt}}p{#1}}
\newcolumntype{R}[1]{>{\raggedleft\arraybackslash}p{#1}}
\newcolumntype{L}[1]{>{\raggedright\arraybackslash}p{#1}}
\begin{document}\sloppy
\begin{frontmatter}
\title{Modelling the pulsatile flow rate and pressure response of a roller-type peristaltic pump}
\author[1]{Michael P. McIntyre}
\author[2]{George van Schoor}
\author[1]{Kenneth R. Uren\corref{cor1}}
\ead{Kenny.uren@nwu.ac.za}
\cortext[cor1]{Corresponding author}
\author[3]{Cornelius P. Kloppers}

\address[1]{School of Electrical, Electronic and Computer Engineering, Faculty of Engineering, North-West University, 11 Hoffman street, Potchefstroom, 2531}
\address[2]{Unit for Energy and Technology Systems, Faculty of Engineering, North-West University, 11 Hoffman street, Potchefstroom, 2531}
\address[3]{School of Mechanical Engineering, Faculty of Engineering, North-West University, 11 Hoffman street, Potchefstroom, 2531}

\begin{abstract}
The unique working mechanics of roller-type peristaltic pumps have allowed their applications to span a wide variety of sectors and industries. The roller-type pump's accurate dosing and hydrostatic capabilities can theoretically allow for the pump to be used for hydraulic actuation (as an electro-hydrostatic actuator) for low pressure applications. This however requires accurate control of the peristaltic pump and its flow rate. The associated pressure pulsations will also have an impact on the pump's selection criteria. Accurate control of roller-type peristaltic pumps commonly rely on flow-meters, which increases the cost of the pump and can complicate control strategies. Current modelling approaches either do not rely on first principle modelling and require expensive simulation software, or do not apply for larger Reynolds numbers at larger flow rates. The most applicable model focusses on the flow rate for each roller individually. This implies that the model requires alterations in order to accommodate pumps with varying numbers of rollers. This paper presents an alternative modelling methodology towards the volume flow rate, pulsatile flow rate qualities, and pressure pulsations commonly found on peristaltic pumps. The model instead focusses on the flow rate at the inlet and the outlet of the pump, rather than on each individual roller. This model is highly scalable and allows for varying number of rollers. The model is validated using a 3D printed peristaltic pump and pulsatile flow rate test bench. The pump is capable of accommodating roller housings with varying numbers of rollers (3 or 2) in order to validate the model.
\end{abstract}


\end{frontmatter}

\section{Introduction}
Roller-type peristaltic pumps (also referred to as roller pumps) generate flow by deforming a selected process tube in a manner that resembles peristalsis in living organisms. This unique working mechanism allows the pumps to isolate the fluid being transported from the mechanical parts of the pump. This isolation reduces contamination to such a degree that these pumps are commonly used as blood pumps found in heart-lung bypass machines used for cardiopulmonary bypass surgery~\cite{Passaroni2015, Mejak2000}. Additionally, the mechanism on which these pumps work can allow for full occlusion of the process tube, essentially sealing the tube at the site of occlusion. Full occlusion can allow for highly accurate dosing applications by means of maintaining a hydrostatic pressure differential between the inlet and outlet of the pump~\cite{peristaltic_pumps_review}. This hydrostatic pressure differential is essential in dosing applications, but can also serve the utilisation of the peristaltic pump as an electro-hydrostatic actuator~(EHA) for low pressure applications. The ability of the pump to handle numerous fluids also extends to gasses, which might prove useful in other applications such as modular soft robotics~\cite{shepherd2011multigait, rus2015}. 

Actuation methods of robotic systems are still considered to be one of the greatest challenges in automation, especially with regards to human-friendly robotics. These human-friendly robotics also extend to rehabilitation devices such as prosthetics and orthesis. Ideally, these actuators would be required to produce large torques at low velocity, have a high power to mass ratio, be highly integratable, and generate smooth 'human-like' motions~\cite{alfayad2011high,alfayad2011high2}. Peristaltic pumps, specifically linear peristaltic pumps named `solid-state electroactive smart material actuators' are stated to  possibly impact various industries in a positive manner. This includes the use thereof as an EHA, however, the extent to which these actuators can be up-scaled is not clear as the maximum flow rate of the reviewed pumps is indicated to be 422~mL/min~\cite{sideris2020pumps}. Rotary peristaltic pumps, such as roller pumps, tend to have larger flow rates than that of their linear counterparts, and can be applicable for faster actuation requirements that require larger flow rates.

The use of a roller pump as an EHA requires accurate control of the fluid passing through the pump for accurate actuation. Generally, roller pumps rely on flow-meters for accurate control. These flow-meters are not only an additional cost to the pump assembly, but also require constant calibration due to manufacturing deviation of the pump and replaceable tubing~\cite{peristaltic_pumps_review}. A model for the flow rate and pressure pulsations can reduce cost and aid control methods when the manufacturing process remains optimally constant. Literature in the public domain, however, is scarce in comparison to the vast number of published roller pump patents. Older modelling literature on roller pumps assumed infinite sinusoidal wave-trains with low Reynolds numbers, which closely resemble natural occurring peristalsis~\cite{weinberg1971experimental,latham1966fluid}. These modelling techniques are not effective when modelling larger flow rates, yet remain useful in detecting unique attributes of viscous flow in roller pumps, such as trapping~\cite{SALAHUDDIN20208337}. More recently, Moscato et al.~\cite{validated} provided a paper on a validated lumped parameter model which is indicated to be accurate with larger flow rates. Instead of assuming an infinite wave-train, the model has an individual flow rate pulsation as an input for each roller coming into or out of contact with the process tube. The pulsation is determined by differentiating the measured volume that the roller displaces as it engages the tube with respect to the motor angle. This allows the premise for modelling roller pumps with varying numbers of rollers. 

The lumped parameter model provided in~\cite{validated} is, however, developed for a two-roller pump. As roller pumps can be configured to have varying numbers of rollers,  the need for a generalised model exits. This model will therefore allow scalability in terms of varying number of rollers. This paper provides a first principles modelling approach for peristaltic pumps with varying numbers of rollers. An alternative model to that provided in~\cite{validated} is discussed in depth with regards to both the modelled flow rate and lumped parameter model. The alternative approach allows for roller pumps to be modelled with varying numbers of rollers by changing the frequency of the controlled sources, rather than changing the model itself.

This paper is outlined as follows: Section 2 discusses the modelling approach and how it varies from current literature. Section 2 outlines the lumped parameter model, flow rates, and frequencies of the controlled sources. Section 3 provides details pertaining to the peristaltic pump used to validate the model provided in this paper. Section 4 details the validation tests and the test bench. The results along with their discussion are provided in Section 5 with the conclusion provided in Section 7.

\section{Modelling approach}
\begin{figure}[]
\hspace*{-0.3cm}
\begin{tabular}{ccc}
\subfloat[]{\includegraphics[width = 0.232\textwidth ,trim={2.2cm 0cm 2.2cm 0cm}, clip]{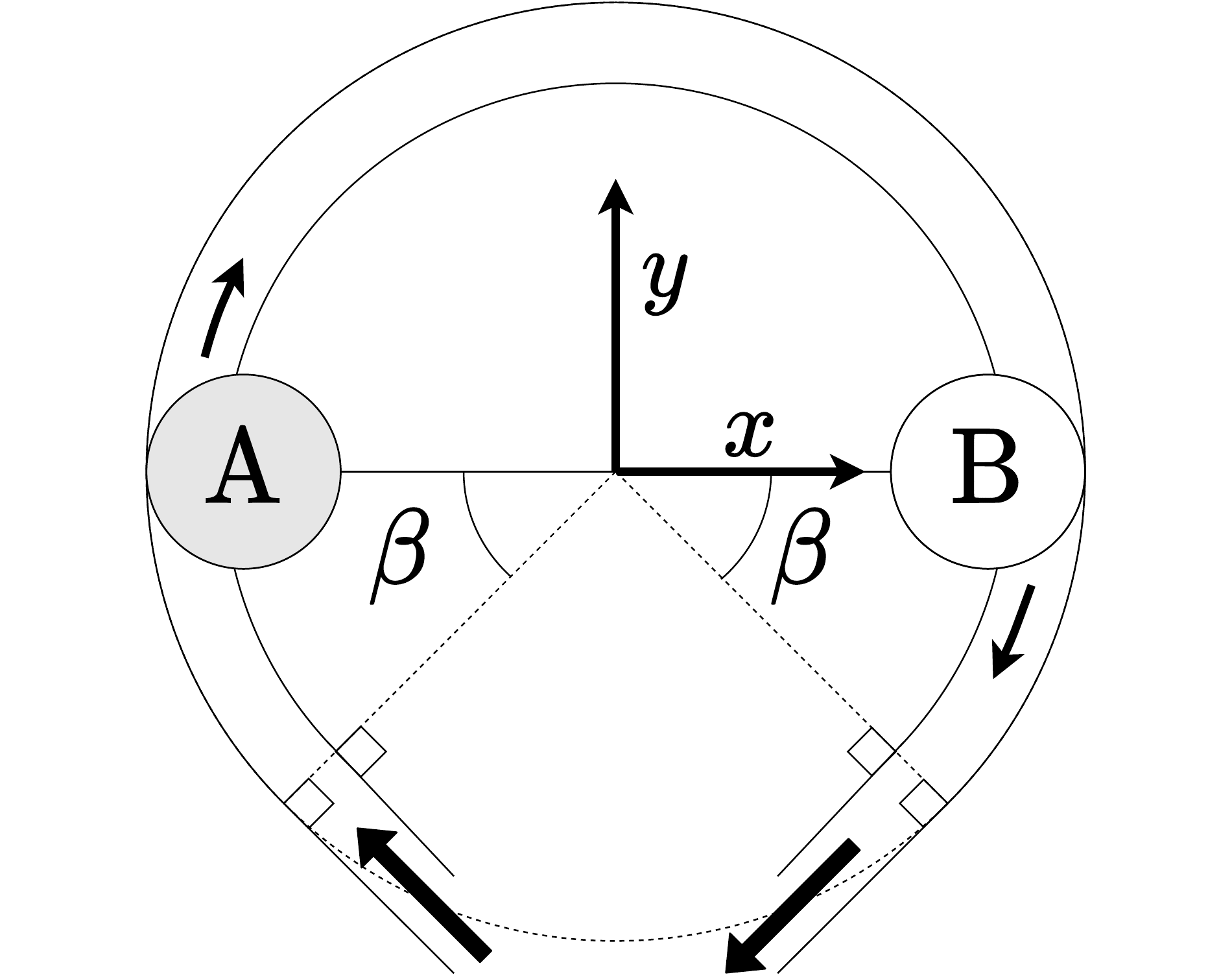}}&
\subfloat[]{\includegraphics[width = 0.232\textwidth ,trim={0.555cm 0cm 0.555cm 0cm}, clip]{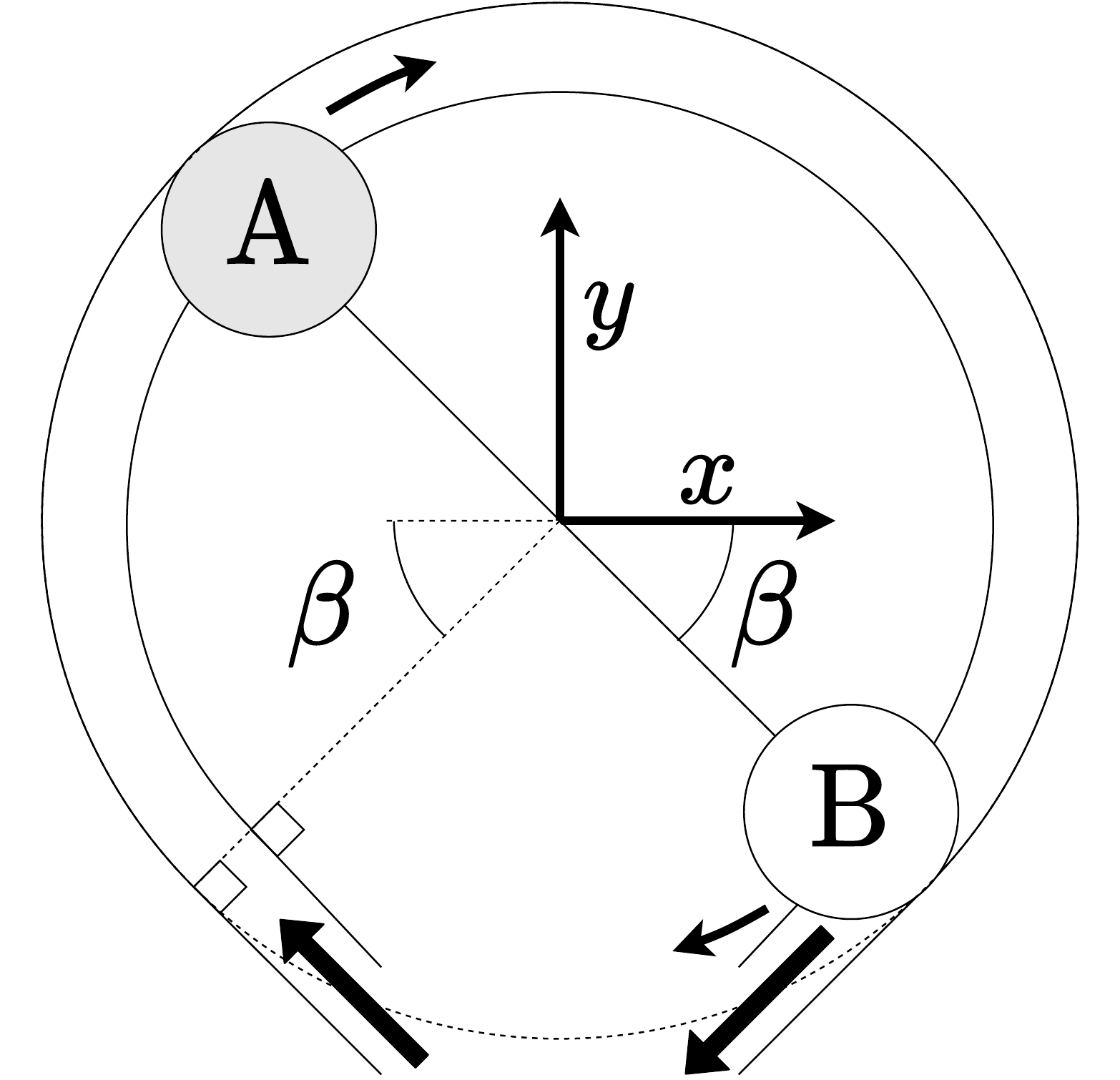}}&
\subfloat[]{\includegraphics[width = 0.232\textwidth]{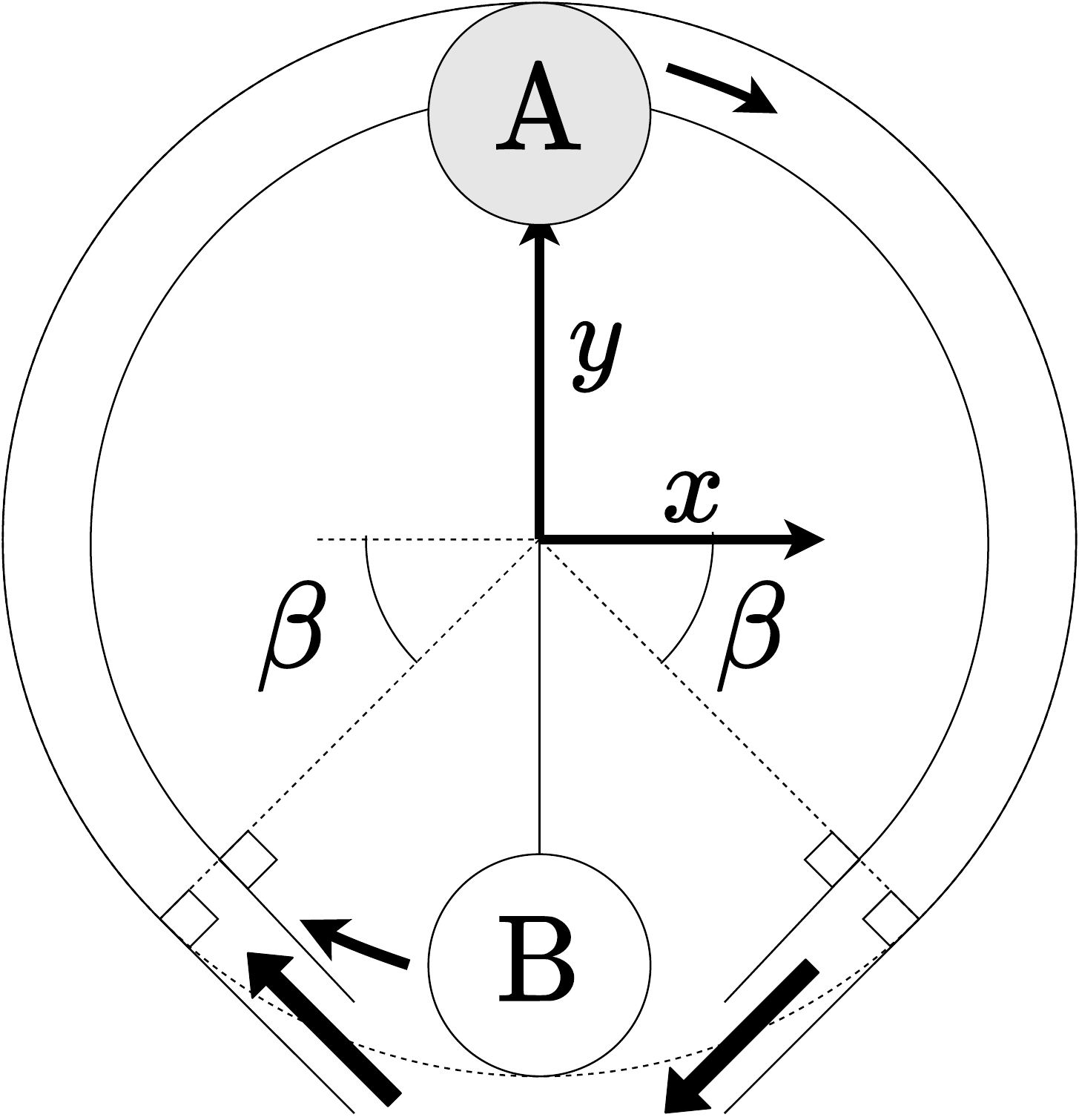}}
\end{tabular}
\caption{Illustration of the sectional view of a symmetrical (around $y$-axis) two-roller pump indicating the working principles by rotating from (a) to (b) and finally (c) via clockwise rotation}\label{roller_pump}
\end{figure}

Fig.~\ref{roller_pump} illustrates the basic working principle of a roller pump with indication of the contact angle $\beta$, discussed later. A roller collapses a process tube while rotating around a central axis at an offset distance. The offset distance (referred to as the roller offset distance) is equivalent to the distance between the central axis of rotation of the pump and the central axis of the roller. This translates the fluid within the tube in the direction of rotation. This differs from linear peristaltic pumps which, in quick summary, imitate peristalsis by moving columns (or segments) of water in succession. Rotary roller-type pumps are, in general, simpler in mechanical design and operation. However, linear peristaltic pumps have somewhat recently found promising capabilities as micro-pumps for accurate dosing and on-chip cooling applications~\cite{sideris2020pumps}.

\floatsetup[figure]{style=plain,subcapbesideposition=top}
This paper focusses specifically on roller pumps with a backplate, which is effectively the flat part of the roller raceway, against which the tube is pressed by the roller. This ensures that full occlusion occurs, which is important when determining the assumptions of the model.
\begin{figure*}[t]
\centering
\begin{tabular}{c}
\subfloat[]{\includegraphics[width = 0.85\textwidth]{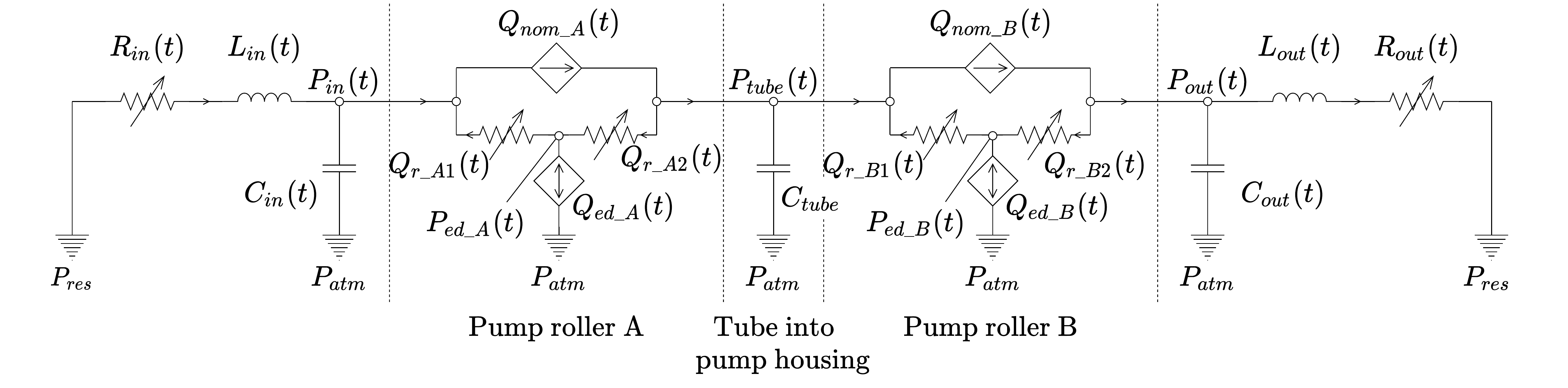}\label{lpm1}}\\
\subfloat[]{\hspace{1cm}\includegraphics[width = 0.85\textwidth]{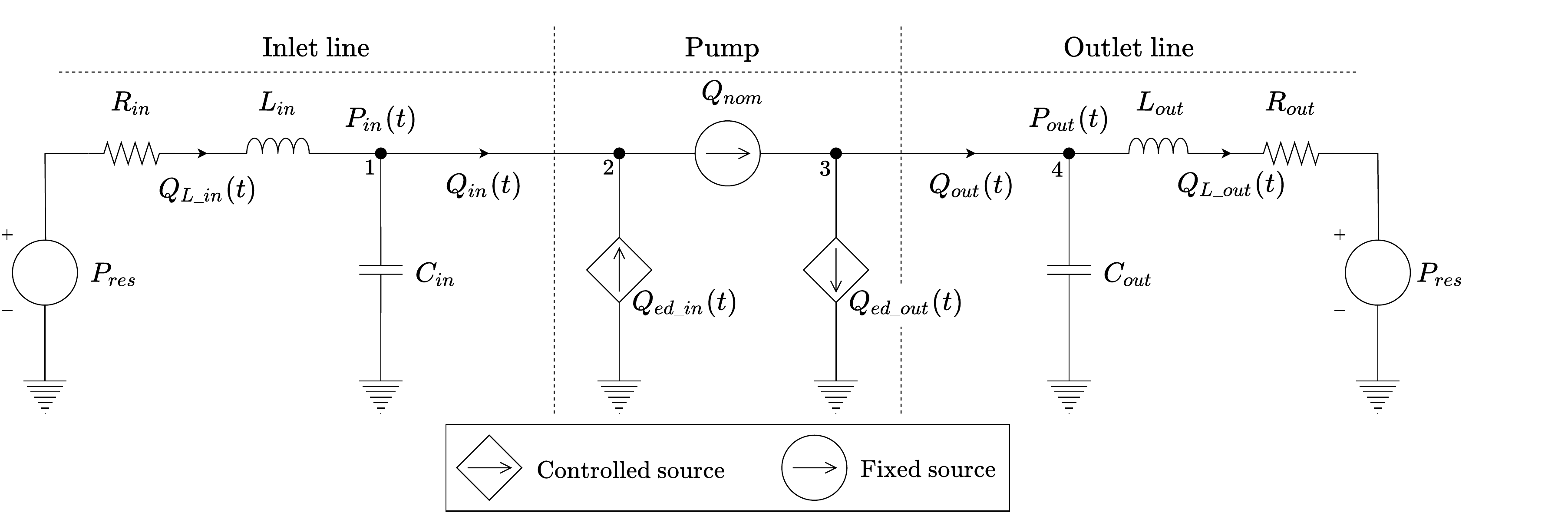}\label{lpm2}}
\end{tabular}
\caption{Lumped parameter models for roller-type peristlatic pumps: (a) Reference (roller-centric) approach from~\cite{validated}, (b) proposed (flow-centric) approach}\label{lpm}
\end{figure*}
Fig.~\ref{lpm} indicates two separate lumped parameter models. Both models are electrical analogues of the actual fluid systems, with the direction of the sources indicated by directional arrows. The lumped parameter model proposed in this paper is developed with the model provided in~\cite{validated} as the initial reference (Fig.~\ref{lpm1}). The referenced model focusses on the pressure at each individual roller by creating a subsystem for each roller. These subsystems are connected to the inlet and outlet lines in series, however, in the subsystem itself parallel circuits are used. It is assumed that for the model to apply to pumps with more rollers, additional sub-systems need to be added to the lumped parameter model. The referenced model is therefore referred to as the roller-centric model. The model provided in this paper (Fig.~\ref{lpm2}) simplifies the referenced model by omitting the subsystems entirely. Instead, focus is placed on the inlet and the outlet of the pump, with the flow rate pulsations of the rollers used as controlled sources. The model provided in this paper is therefore referred to as the flow-centric model. As total occlusion is assumed, the maximum velocity through the pump ($Q_{nom}$) is selected to be a DC current source in the electrical analogue. Representing the nominal flow rate as a DC source is justified as positive displacement pumps are commonly modelled as current sources~\cite{karassik2000pump}. The flow rate pulsations ($Q_{ed}$) associated with the rollers are inserted as variable flow sources directly before and after the the DC current source. As both models are electrical analogues of hydraulic circuits: the resistance $R$, capacitance $C$, and inductance $L$ are equivalent to the effective fluid friction, compliance, and inertia respectively. This also implies that pressure is analogous to the voltage, and the volume flow rate analogous to the current.

The roller induced flow rate at the inlet of the pump ($Q_{ed\_in}$) is, for the proposed model, always positive while that of the outlet ($Q_{ed\_out}$) is always negative. This allows the roller induced flow rate to always work in an opposing direction to normal flow. This differs from the roller-centric model where the roller induced flow rates ($Q_{ed\_A}$ and $Q_{ed\_B}$) alternate between positive and negative values for each roller. The $P_{res}$ pressure sources in the model normalise the pressures of the inlet and outlet to that of the ambient pressure caused by the water level in the reservoir tank. This takes into account the varying common ground potentials ($P_{res}$ and $P_{atm}$) of the roller-centric model. The capacitance value of the flow-centric model should therefore be calculated/measured with reference to atmospheric pressure. It should also be noted that the current source `blocks' (both variable and static) provided in the Simscape library (used to simulate the lumped parameter model) do not allow current to flow through the block other than that which the block is set to distribute. The following equations relate instantaneous inlet and outlet pressures and flow rates. Applying Kirchhoff's current law for node 1 in Fig~\ref{lpm2} gives:
\begin{linenomath*}\begin{equation}
Q_{L\_in}(t) - Q_{in}(t) = C_{in}\dfrac{\text{d}P_{in}}{\text{d}t},
\end{equation}\end{linenomath*}
while the same law for node 2 gives:
\begin{linenomath*}\begin{equation}
Q_{in}(t) = Q_{nom} - Q_{ed\_in}(t).
\end{equation}\end{linenomath*}
Similarly, applying Kirchhoff's current law to node 4 gives:
\begin{linenomath*}\begin{equation}
Q_{out}(t) - Q_{L\_out}(t) = C_{out}\dfrac{\text{d}P_{out}}{\text{d}t}, 
\end{equation}\end{linenomath*}
while node 3 gives:
\begin{linenomath*}\begin{equation}
Q_{out}(t) = Q_{nom} + Q_{ed\_out}(t).
\end{equation}\end{linenomath*}
Applying Kirchoff's voltage law to the inlet line connected to node 1 (for $Q_{L\_in}(t)$) gives:
\begin{linenomath*}\begin{equation}
P_{res} - P_{in}(t) = R_{in}\cdot Q_{L\_in}(t) + L_{in}\dfrac{\text{d}Q_{L\_in}}{\text{d}t}.
\end{equation}\end{linenomath*}
Applying the same law for the outlet line connected to node 4 (for $Q_{L\_out}(t)$) gives:
\begin{linenomath*}\begin{equation}
P_{out}(t) - P_{res} = R_{out}\cdot Q_{L\_out}(t) + L_{out}\dfrac{\text{d}Q_{L\_out}}{\text{d}t}.
\end{equation}\end{linenomath*}

\begin{figure}[]
\centering
\begin{tabular}{cc}
\subfloat[P1 -- Phase 1]{\hspace{-0.4cm}\includegraphics[width = 0.37\textwidth , trim={5.2cm 11.5cm 6.2cm 8.9cm},clip]{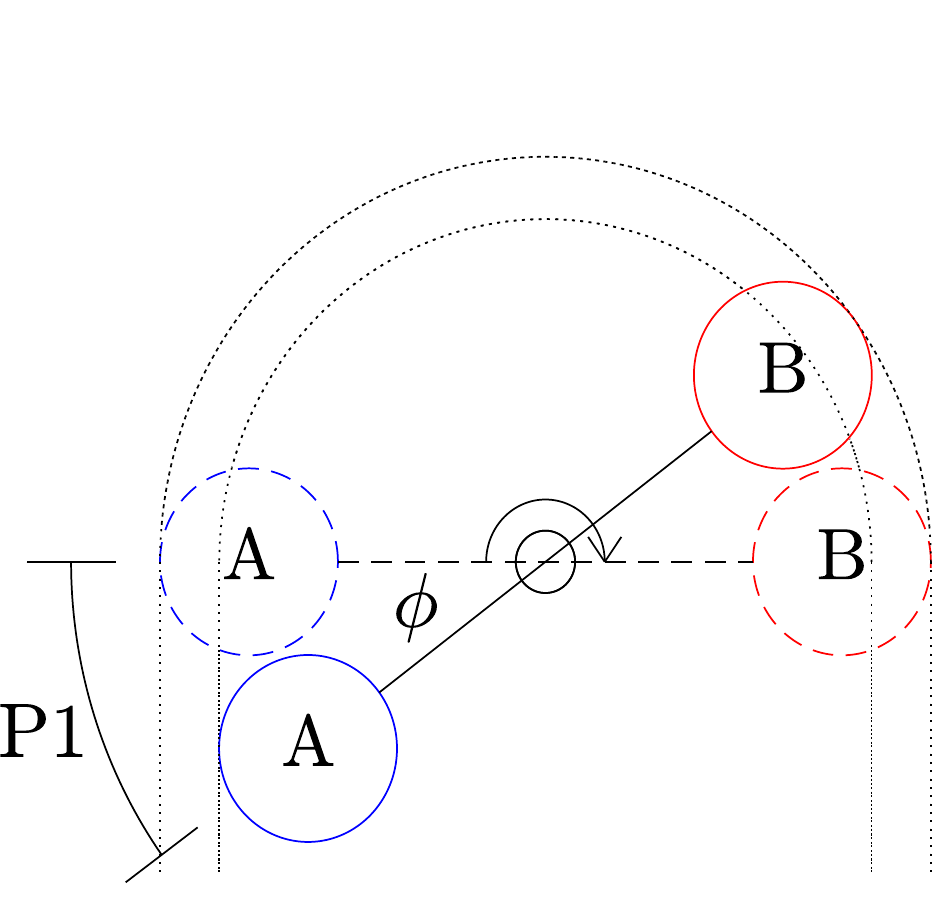}\label{phase1}}&
\subfloat[P2 -- Phase 2]{\includegraphics[width = 0.408\textwidth ,trim={5.5cm 11.5cm 5cm 8.85cm},clip]{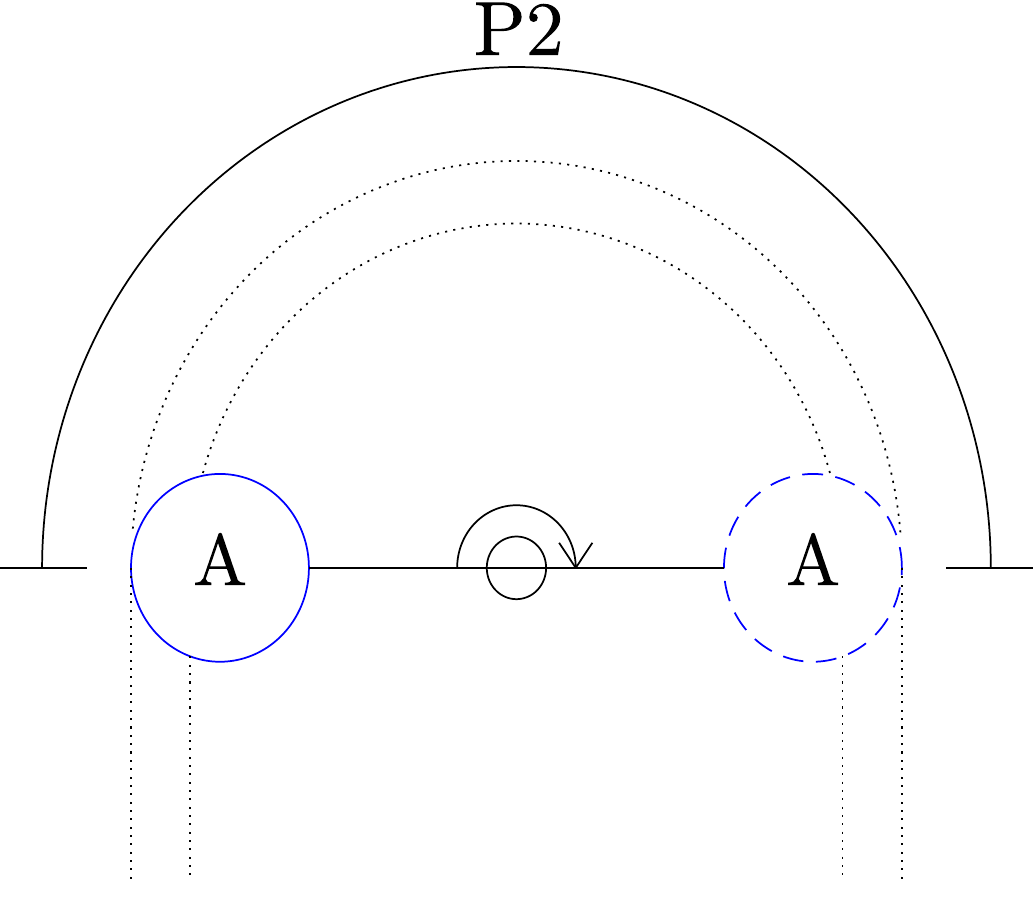}\label{phase2}}\\
\subfloat[P3 -- Phase 3]{\hspace{0.4cm}\includegraphics[width = 0.37\textwidth ,trim={6.7cm 10.0cm 4.7cm 10cm},clip]{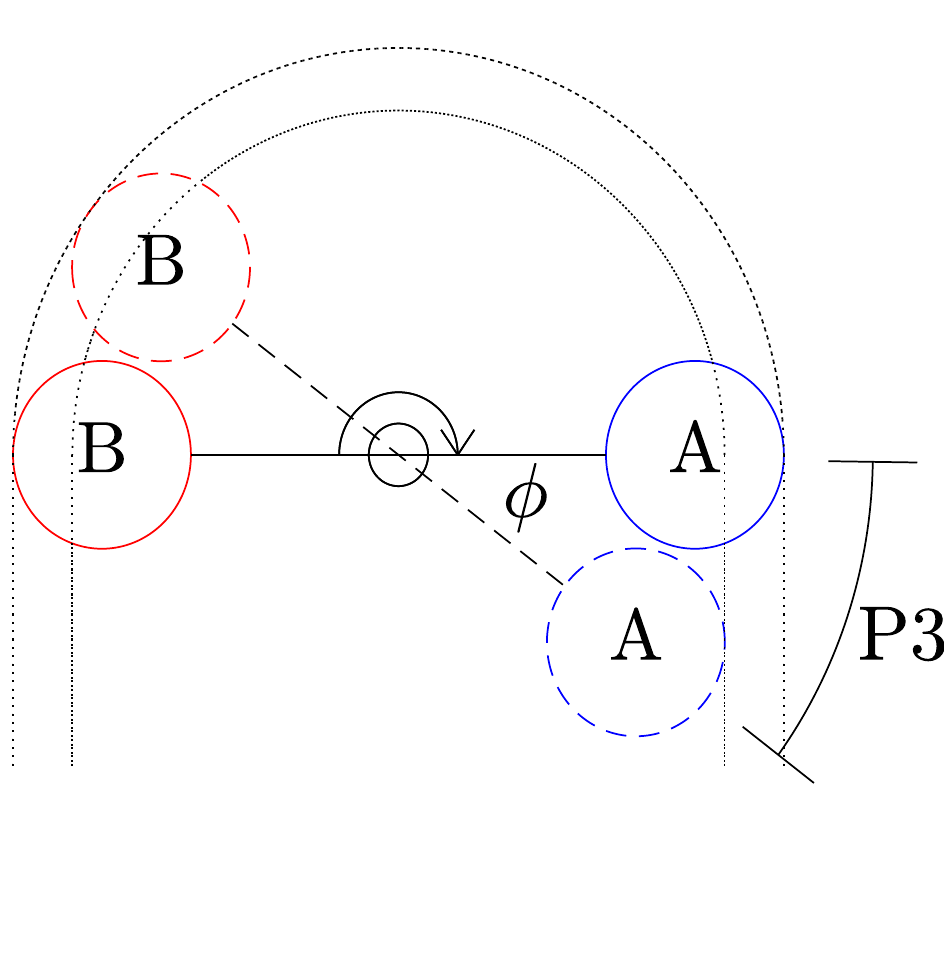}\label{phase3}}&
\subfloat[P4 -- Phase 4]{\includegraphics[width = 0.408\textwidth ,trim={5.5cm 10.0cm 5cm 10cm},clip]{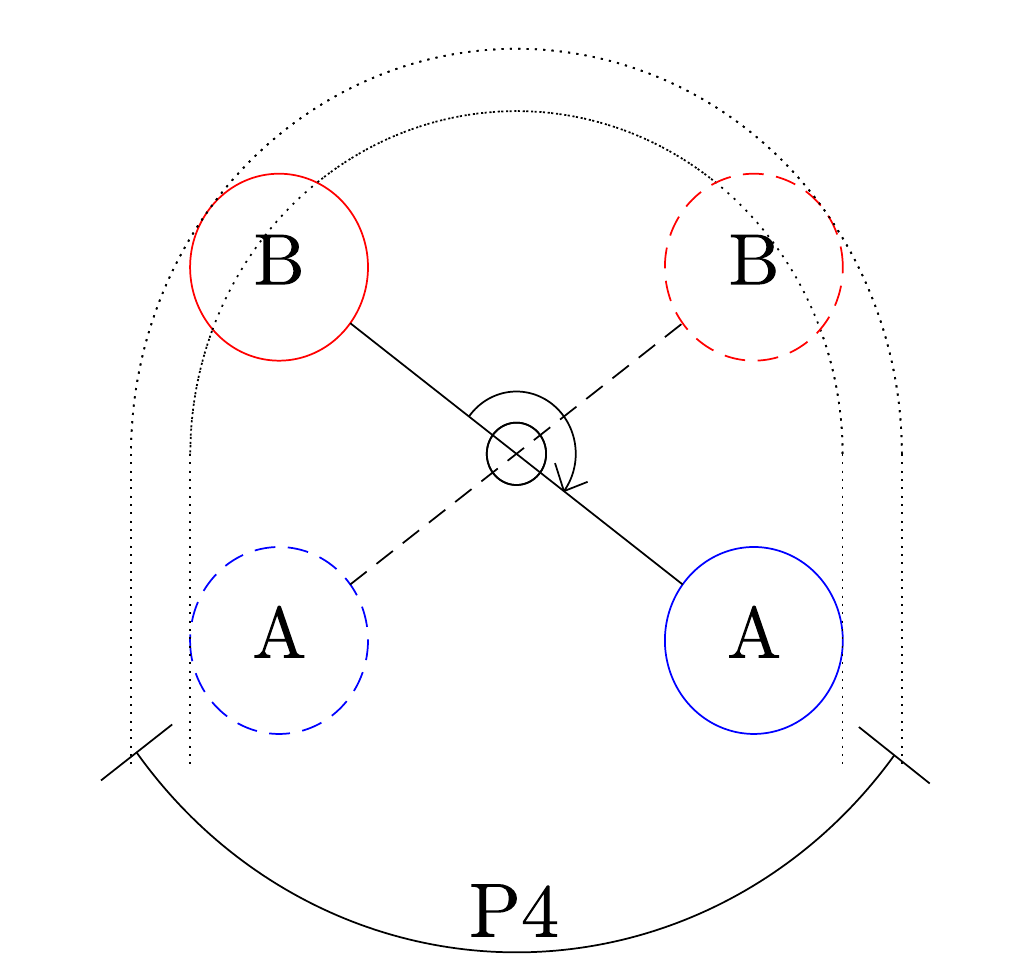}\label{phase4}}\\
\end{tabular}
\caption{Illustration of the sectional view of a two-roller peristaltic pump indicating the roller positions pertaining to the pump phases, $\beta = 0~\degree$}\label{phases}
\end{figure}

\begin{figure}[t]
\centering
\begin{tabular}{c}
\subfloat[]{\hspace{-0.2cm}\includegraphics[width = 0.71\textwidth]{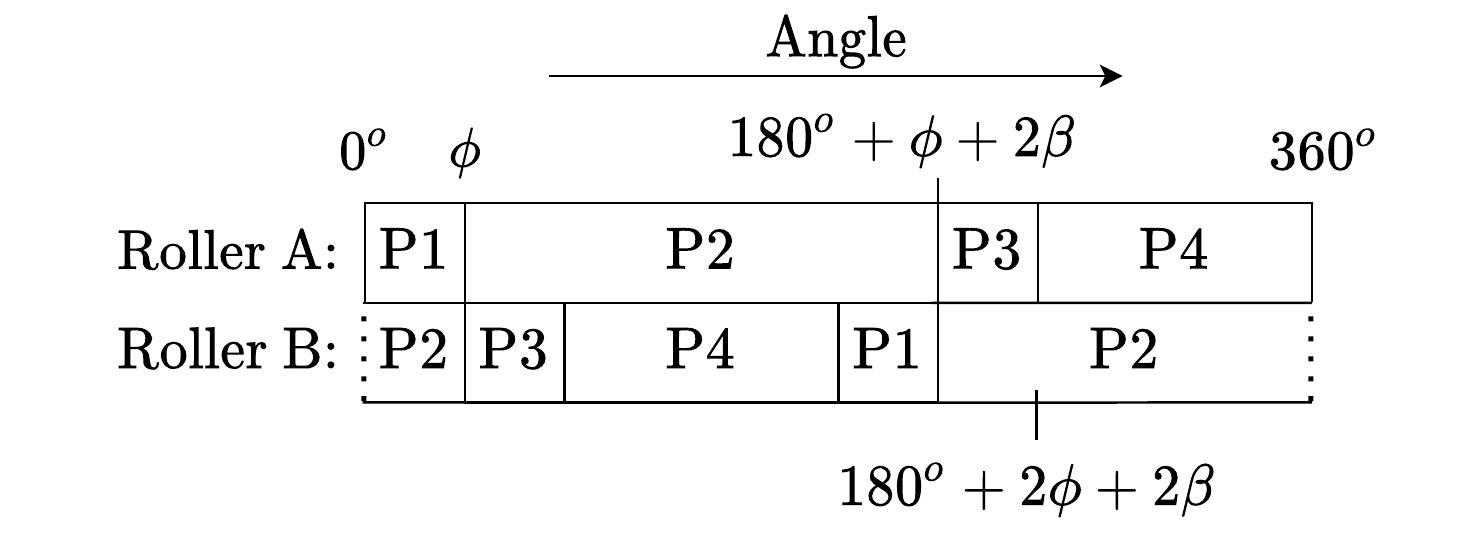}\label{roller_centric_pd}}\\
\subfloat[]{\includegraphics[width = 0.68827\textwidth, trim = {2.2cm 0cm 0cm 0cm}, clip]{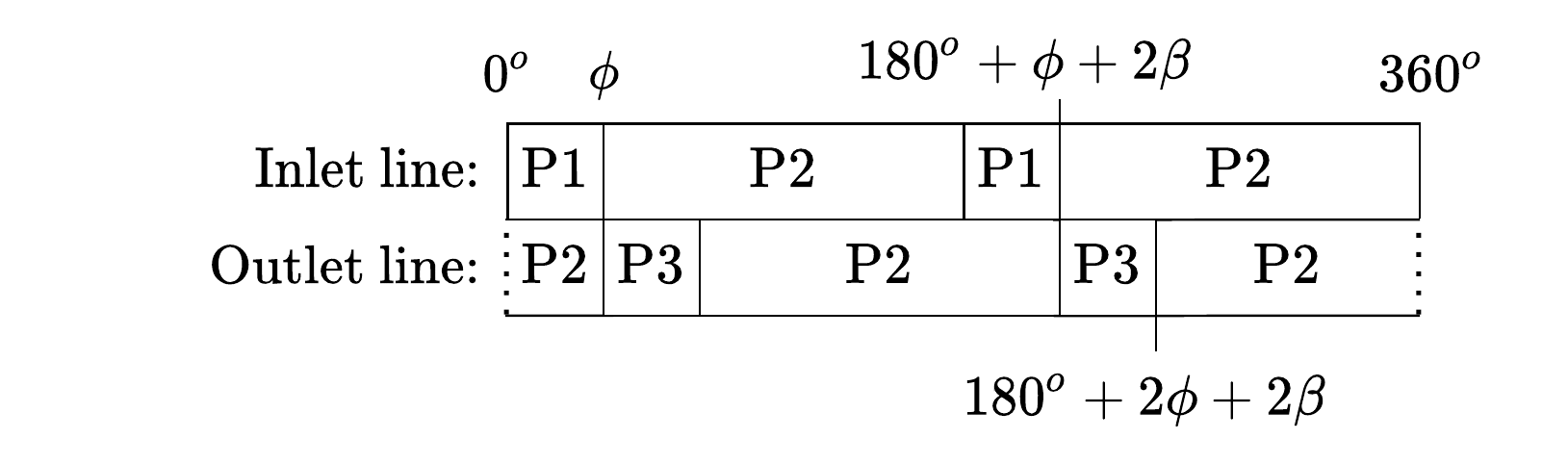}\label{flow_cntric_pd}}
\end{tabular}
\caption{Illustration of the (a) roller-centric and (b) flow-centric phases at various motor angles}\label{phase_diagrams}
\end{figure}

\begin{figure}[]
\centering
\includegraphics[width = 0.8\textwidth]{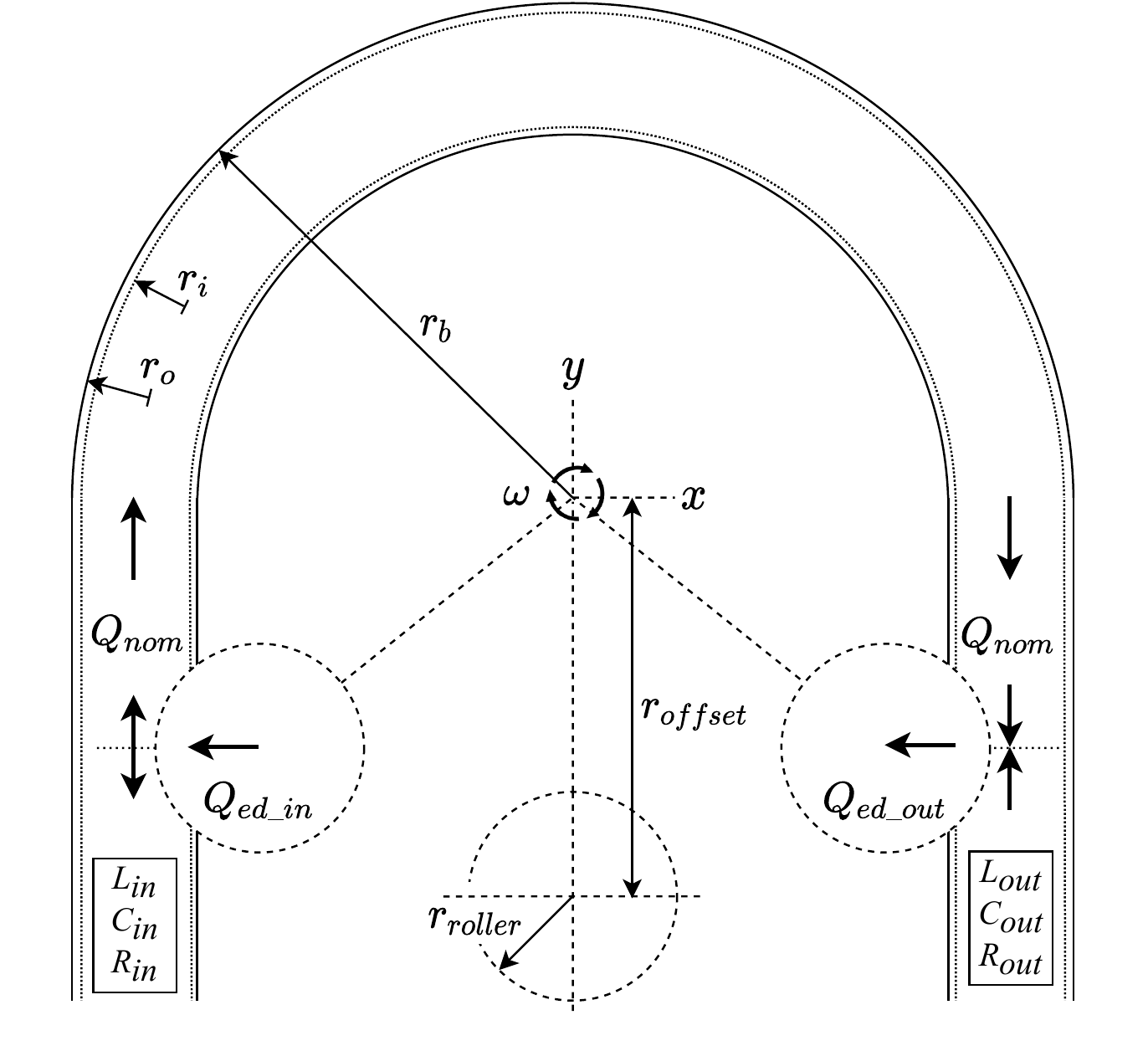}
\caption{Illustration of the sectional view of a roller pump indicating the modelling variables related to the geometry of the pump, as well as the induced flow rates, $\beta = 0~\degree$}\label{variables}
\end{figure}

Fig.~\ref{phases} illustrates the various phases of the roller outlined in~\cite{validated}. The various phases can be labelled as: Phase 1 - engaging phase, phase 2 - engaged phase, phase 3 - disengaging phase, phase 4 - disengaged phase. Fig.~\ref{phase_diagrams} illustrates the individual phases associated to the roller position in chronological order for both the roller-centric and flow-centric model. For the flow-centric model, the phases are associated with the following flow rates: Phase 1 - $Q_{ed\_in}$, phase 2 - $Q_{nom}$, phase 3 - $Q_{ed\_out}$, with phase 4 delivering no flow. As phase 4 delivers no flow, the phase is not indicated in Fig.~\ref{flow_cntric_pd} as focus is not placed on the roller, but rather the inlet and outlet lines. It should be noted that as a roller engages or disengages that the opposing roller is in the engaged phase, thus $Q_{nom}$ is always present. Additionally, the contact angle $\beta$ is also taken into account, allowing pumps that are fully engaged for ranges differing from 180~\degree ~to be modelled.  The contact angle $\beta$ (as indicated in Fig.~\ref{roller_pump}) describes the span of the backplate with reference to the \textit{x}-axis of the pump. Note that $\beta$ is equivalent to the absolute value of this angle. This angle plays a role in the calculation of the phase of the pulsations with respect to the inlet and outlet. Fig.~\ref{variables} illustrates the various modelling variables as well as the roller induced flows and nominal flow with a sectional view of a roller pump. 

The following model assumptions apply:

\begin{enumerate}
\item The tube conforms perfectly to the backplate walls.
\item The pump is symmetrical around the \textit{y}-axis.
\item The rollers are placed symmetrically around the axis of rotation.
\item No leakage occurs across the site off occlusion when a roller is fully engaged (full occlusion) across the entire span of engagement.
\item There is always one roller fully engaging the tube (required for hydrostatic stability).
\item The rollers and pump components are completely rigid.
\item Temperature remains constant during operation.
\end{enumerate}

For simplicity, the rollers are assumed to slide over the tube with negligible friction. This allows for reference of the axis of rotation to apply specifically to the central shaft of the pump. The pulsatile flow rate comprises both the nominal flow rate ($Q_{nom}$) and roller induced flow rate ($Q_{ed}$). With the assumption that the pump is symmetrical around the \textit{y}-axis, it can be said that the roller induced flow rate at the inlet and outlet are equal in magnitude, but reversed in direction. This is discussed in more detail in the following sub-section labelled `Pulsatile flow rate'. Furthermore, the assumption of full occlusion allows for an additional assumption, namely: that the complete magnitude of the roller induced flow rate opposes normal flow at both the inlet and outlet as opposed to only half the magnitude. The values of the parameters ($R$, $L$, $C$) in the lumped parameter model are equivalent to the effective parameter values. This implies that all resistance, inertial, and compliance values of the system are taken into account on both the inlet and outlet.

The numerical modelling of the flow rate is done in MathWorks' MATLAB, with the lumped parameter model compiled and solved in Simulink/Simscape. The simulation of the lumped parameter model is solved using the ode15s (stiff) solver with a maximum step time of $h = 0.001$~s. The key elements of the Simulink model are provided by the Simscape library, such as resistors, capacitors, and inductors. The numerical input values are solved and stored in MATLAB's workspace and are imported into the simulated model via Simulink components. The controlled sources of the numerical flow rate values are solved using the same time increment $h$ for continuity. The Simulink simulation of the lumped parameter model as well as the MATLAB files used to determine the controlled sources and define the model parameters have been placed in a GITHUB repository at the following link: \url{https://github.com/MPMcIntyre/Roller-pump-model-and-simulation}. The simulation and code have been made available to fellow researchers to suggest possible improvements and critiques to the authors in order to improve the model accuracy.

\subsection{Pulsatile flow rate}
As stated previously, the pulsatile flow rate can be modelled taking both the nominal flow rate and the roller induced flow rate into account. The nominal flow rate is a static value determined using the motor's rotational speed and pump dimensions. The roller induced flow rate is a vector of zeros with interpolated values of the differentiated volume polynomial inserted at the engaging/disengaging time intervals as indicated in Fig.~\ref{RIF}. For clarification, a vector refers to a $1\times n$ or $n\times 1$ matrix in the MATLAB environment. The variable flow rate is thus correlated to the time vector that ranges from 0 to $t_{sim}$ in increments of $h$, with $t_{sim}$ the total simulation time. The methods for determining the nominal, roller induced, and average flow rate are discussed next.

\subsubsection{Nominal flow rate}
 The nominal flow rate describes the idealistic flow through the pump with negligible roller induced flow rates and roller volume displacements. The nominal flow rate also describes the maximum flow rate achieved by the pump while a roller is not engaging (for the inlet flow rate) or disengaging (for the outlet flow rate) the tube. The model provided in~\cite{validated} determines the nominal flow rate using the cross sectional area of the tube as the normal surface. The radial distance, which is used to determine the flow rate as a function of the motor speed ($\omega$), is determined from the offset radius rather than the centre of the tube. The nominal flow rate calculation with regards to the assumptions made in this paper is improved by calculating the nominal flow rate with reference to the centre of the tube as:
 
\begin{linenomath*}\begin{equation}
Q_{nom} = \pi r_i^2 \cdot \omega(r_b - r_o)
\end{equation}\end{linenomath*}

\subsubsection{Roller induced flow rate}
\begin{figure}[!]
\centering
\begin{tabular}{c}
\subfloat[]{\includegraphics[width = 0.8\textwidth , trim = {3.3cm 12.4cm 4cm 12.4cm}, clip]{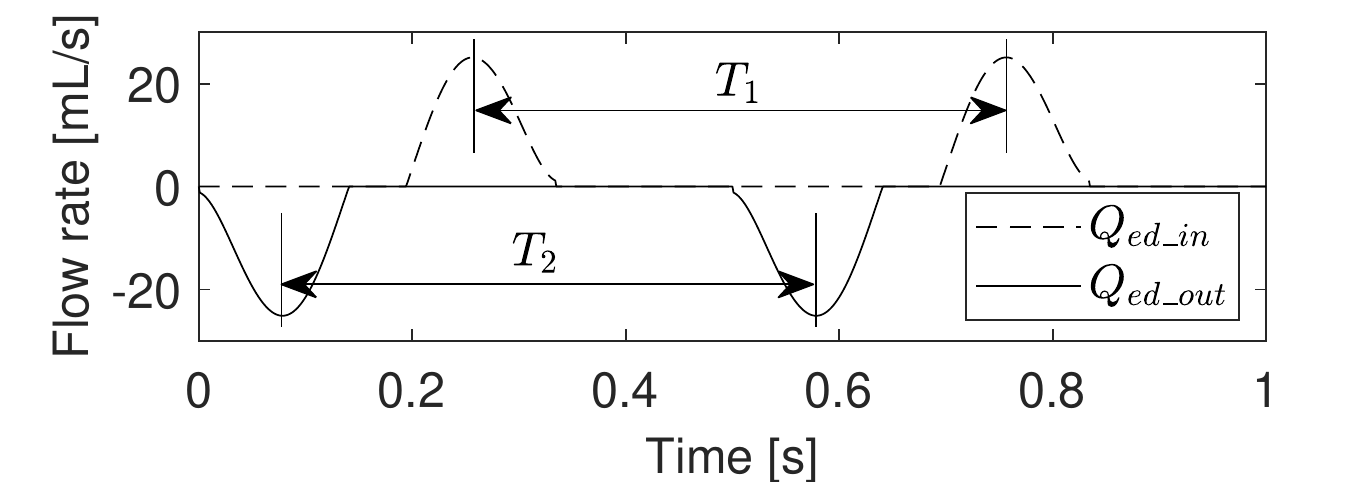}}\\
\subfloat[]{\includegraphics[width = 0.8\textwidth , trim = {3.3cm 12.4cm 4cm 12.4cm}, clip]{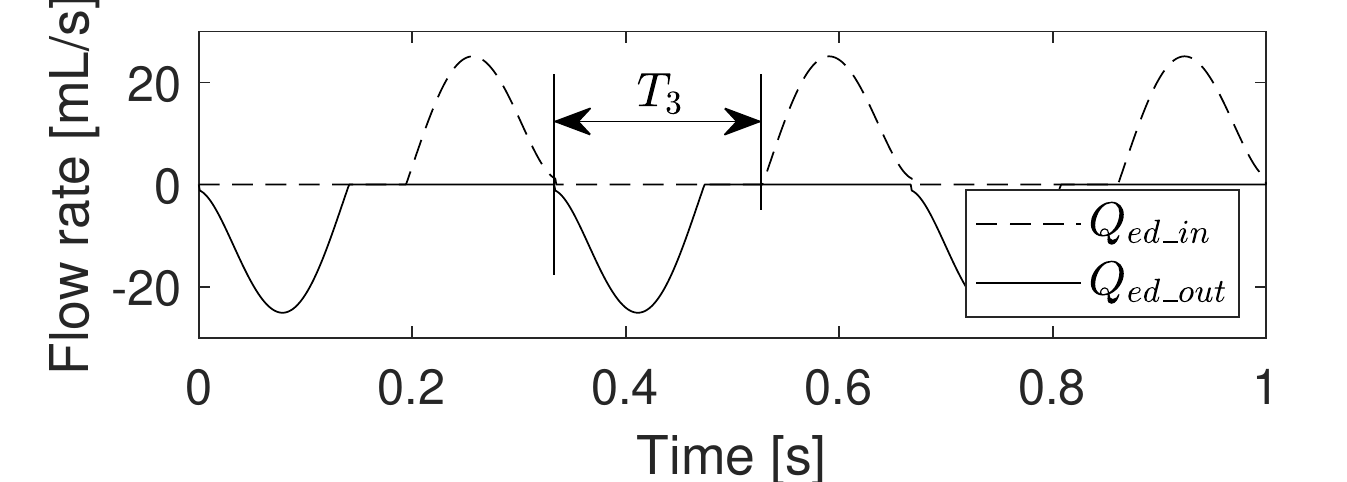}}
\end{tabular}
\caption{Roller induced flow rate vectors for the inlet and outlet for (a) a two-roller and (b) a three-roller pump with a motor speed of 60 r/min, $\beta = 30~\degree$}\label{RIF}
\end{figure}

The roller induced flow rate can be calculated by means of differentiating the roller volume displacement in terms of time or motor angle. To do so, the average roller volume displacement curve is cubically interpolated as a 5th order polynomial as a function of the motor angle $\theta$. With $\theta = t\cdot \omega$, the roller induced flow rate can  be calculated by differentiating the roller volume function with respect to time as $Q_{ed}(t) = \dfrac{dV(t\cdot \omega)}{dt}$. The roller induced flow rate function is then evaluated in time increments of $h$ for the total angle of rotation for the engaging roller. The values are then inserted into a vector of zeros at the corresponding time values where the roller engages/disengages the tube as indicated in Fig.~\ref{RIF}. Values of the roller induced flow rate, usually found at the boundaries, which do not comply to the direction (by sign) of the specified flow rate are set equal to zero.

The time values where the evaluated roller induced flow rate function is inserted into the zero vector are dependent on the angle between each roller and the rotational speed. As it is assumed that the rollers are placed symmetrically around the axis of rotation, the angle and time values where the roller engages or disengages can be calculated with the number of rollers on the pump, angle of engagement, contact angle, and rotational speed of the motor. The model assumes that the roller positions begin with a roller at the position where disengagement starts. The vector is then populated with this position as the reference with a constant rotational speed. The times between flow rate peaks, and thus vector insertion points, can be calculated for both the inlet ($T_1$) and outlet ($T_2$) as:

\begin{linenomath*}\begin{equation}
T_1 = T_2 = \dfrac{2\pi}{NU\cdot\omega},
\end{equation}\end{linenomath*}

\noindent with $NU$ as the number of rollers on the pump. The period between the starting times of the outlet and inlet pulses ($T_3$) spans the angle from the disengaging position to the engaging position. Using the contact angle ($\beta$) and angle of engagement ($\phi$) in degrees, the separation period is calculated as:

\begin{linenomath*}\begin{equation}
T_3 = \dfrac{\pi\left( 180-\phi-2\beta\right)}{180\cdot \omega}.
\end{equation}\end{linenomath*}

\subsubsection{Average flow rate}
The average flow rate of the pump can be determined as a function of the volume that the pump displaces per rotation ($V_{rot}$) and the rotational speed of the motor. The volume displacement per rotation can be determined with the nominal volume displacement of the pump ($V_{nom}$) and maximum roller volume displacement value ($V_r$). With the assumption that the tube adheres perfectly to the backplate, $V_{nom}$ can be determined with the given pump dimensions as:
\begin{linenomath*}\begin{equation}
V_{nom}= 2\pi (r_b-r_o)\cdot \pi r_i^2.
\end{equation}\end{linenomath*}
The realistic volume can be determined by taking into account the maximum volume displacement ($V_r$) of the rollers and the number of rollers in the pump. The volume displaced per rotation of the pump can thus be calculated as:

\begin{linenomath*}\begin{equation}
\label{vol}
V_{rot} = V_{nom} - V_r\cdot NU.
\end{equation}\end{linenomath*}

The average flow rate of the pump is then determined as a function of the motor speed with the rotational volume displacement of the pump as:

\begin{linenomath*}\begin{equation}
\label{flow}
Q_{avg} = \dfrac{V_{rot}\cdot \omega}{2\pi}.
\end{equation}\end{linenomath*}

Similarly, the average flow rate for a constant motor speed $\omega$ can be determined by subtracting the roller induced flow rate vector from the nominal flow rate value and determining the average of the resultants so that:

\begin{linenomath*}\begin{equation}
Q_{avg} = Q_{nom} -\int _0 ^{t}{ Q_{ed\_in}(t)}dt.
\end{equation}\end{linenomath*}

The idealistic flow rate that can be expected directly at the inlet and outlet of the pump, without taking into account the dynamic response of the system, can be calculated as:
\begin{linenomath*}\begin{equation}
Q_{in}(t) = Q_{nom} - Q_{ed\_in}(t),
\end{equation}\end{linenomath*}
and
\begin{linenomath*}\begin{equation}
Q_{out}(t) = Q_{nom} + Q_{ed\_out}(t).
\end{equation}\end{linenomath*}

\section{Pump description}
This section provides information regarding the pump used to validate the proposed model in this paper. The pump specifics provided will enable the determination of the model parameters in order to compare the simulated results to the test results. The specifics can be categorised as follows: General pump description, relevant pump dimensions needed for modelling purposes, and the roller volume measurement of the pump. 

\subsection{General}
\begin{figure}[t]
\centering
\includegraphics[width = 0.9\textwidth]{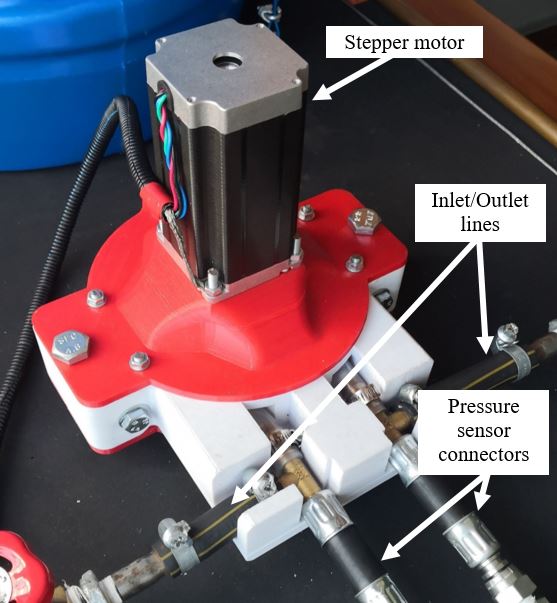}
\caption{Physical peristaltic pump used for model validation}\label{manufactured_pump}
\end{figure}

\begin{figure}
\hspace*{-0.3cm}
\begin{tabular}{cc}
\subfloat[]{\includegraphics[width = 0.38\textwidth, trim = {0cm 0cm 0cm -0.8cm}]{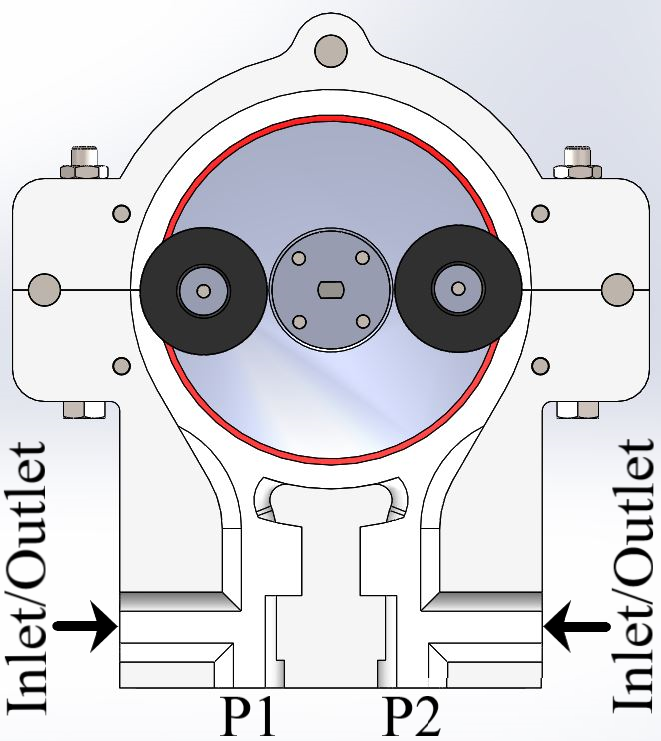}} &
\subfloat[]{\includegraphics[width = 0.455\textwidth, trim = {0cm 0cm 0cm 0.5cm}]{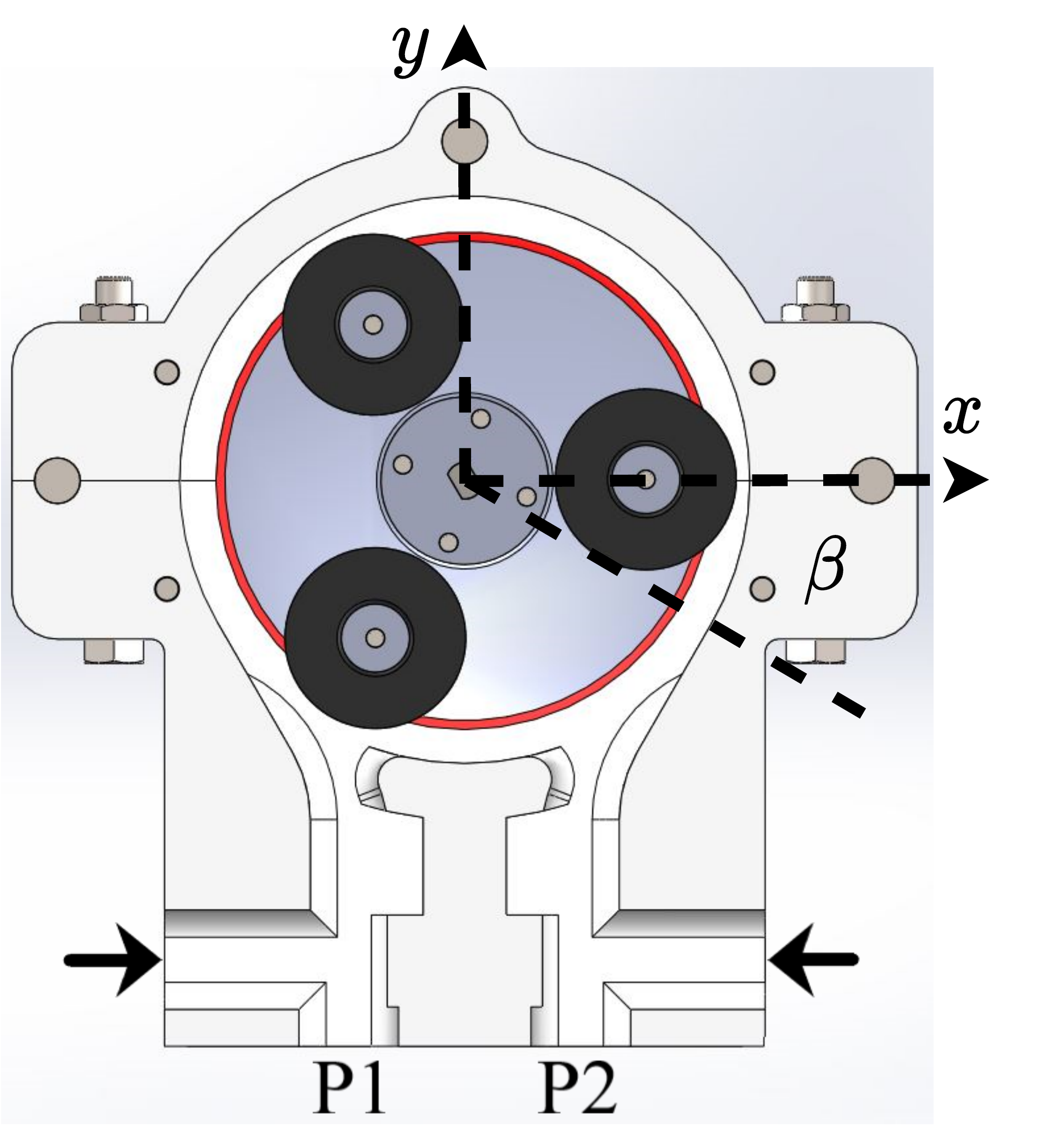}\label{beta_im}}
\end{tabular}
\caption{Cross-sectional rendering of (a) the two-roller and (b) the three-roller configurations of the peristaltic pump with the inlet/outlet indicated by the arrows}\label{pump}
\end{figure}

The physical pump for which the lumped parameter model is developed, is illustrated in Fig.~\ref{manufactured_pump}. The pump was specially developed for the purpose of validating the model~\cite{McIntyre2020}. The mechanical parts were manufactured by means of fused filament fabrication~(FFF) using polyethylene terephthalate~(PETG) on the Prusa MK 2.5 and Prusa MK 3. The pump is and powered by a 3~Nm stepper motor (model: 57BYGH115-003). The pump utilises a single raceway and interchangeable roller housings. This allows for both two-roller and three-roller configurations of the same pump. Fig.~\ref{pump} illustrates this with the pressure sensor outlets as P1 and P2, rollers in black, and roller housing in grey.

\subsection{Pump dimensions}
Table~\ref{dimensions} indicates the relevant dimensions which are used to calculate the theoretical flow rate of the pump with a specified motor speed $\omega$. The entrance and exit to the pump are tangent to the backplate and leave the backplate perpendicular to the specified contact angle. The pump is symmetrical around its \textit{y}-axis, implying that the contact angle for the inlet and outlet are equal. A contact angle of 0~\degree ~specifies a backplate span of 180~\degree , where the pump in this paper has a contact angle of 30~\degree ~and a backplate span of 240~\degree as indicated in Fig.~\ref{pump}. 

\begin{table}
\centering
\caption{Relevant pump dimensions, illustrated in Fig.~\ref{variables}}\label{dimensions}

\begin{tabular}{p{3.cm}x{1.5cm}x{1.15cm}x{1cm}}
\toprule
\textbf{Description:} & \textbf{Symbol:} & \textbf{Value:} & \textbf{Unit:}\\
\midrule
Inner tube radius 	& $r_i$ 		& 5 & mm \\
Outer tube radius 	& $r_o$ 		& 7 & mm \\
Backplate radius 	& $r_b$ 		& 63 & mm \\
Roller radius 		& $r_{roller}$ 	& 20 & mm \\
Roller offset radius& $r_{offset}$ 	& 40 & mm \\
Contact angle	 	& $\beta$ 		& 30 & \degree \\
Max roller volume & $V_r$ 		& 2.06 & mL \\
\bottomrule
\end{tabular}
\end{table}

\subsection{Experimental roller volume displacement measurement}
\begin{figure}
\centering
\includegraphics[width = 0.8\textwidth , trim = {3.3cm 9.3cm 4cm 10cm}, clip]{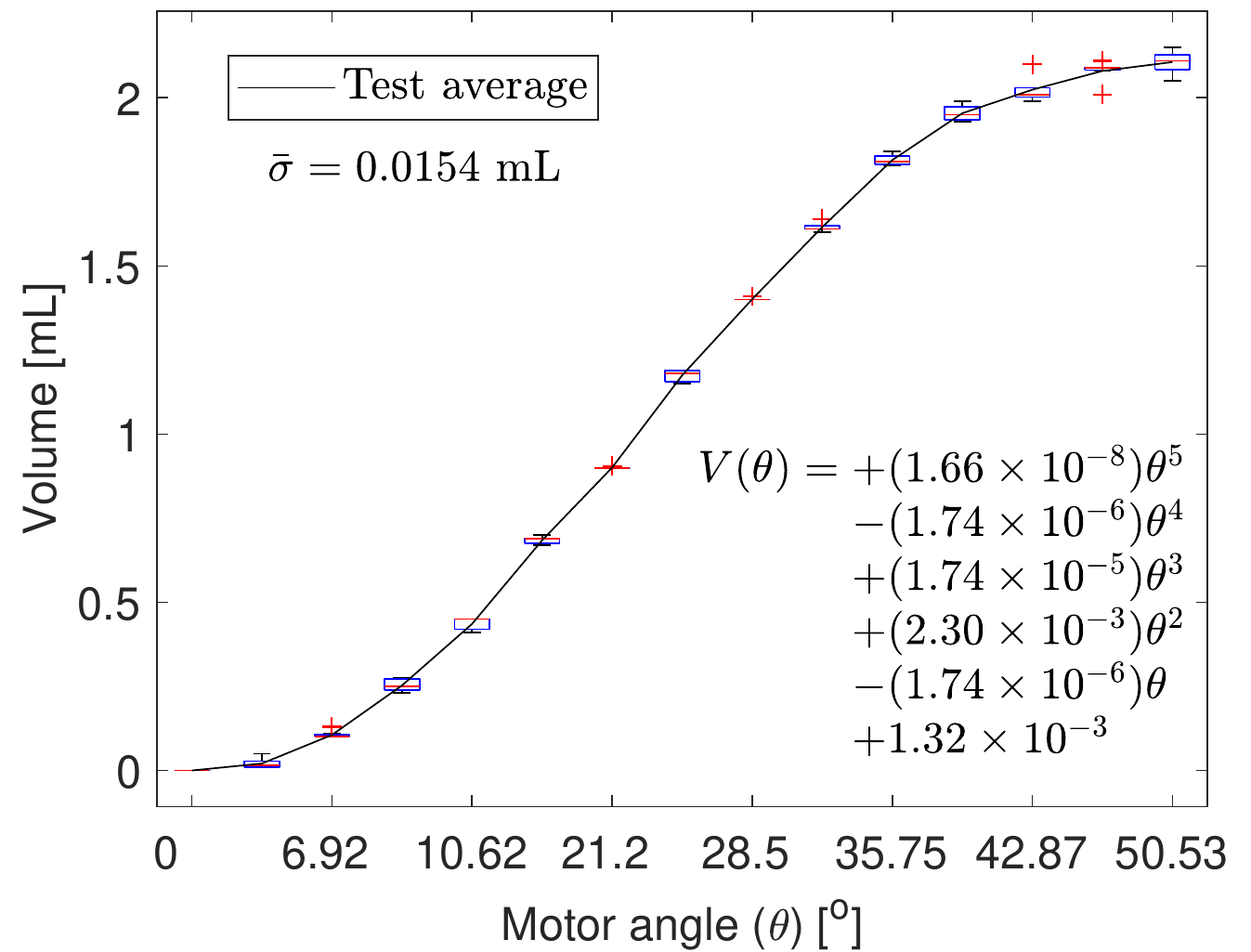}
\caption{Roller volume displacement values over incremental angles for the peristaltic pump with cubically interpolated polynomial function $V(\theta)$}\label{RVD}
\end{figure}

The roller volume displacement for this pump is indicated in Fig.~\ref{RVD}. The roller volume displacement is measured by filling the internal tube segment with water, clamping the tube at the outlet, and removing all but one roller. The motor is incrementally rotated a specified angle (3.63~\degree) from a position where the tube and roller are not in contact to where full occlusion occurs. The volume that the roller displaces as it comes into contact is displaced upward into a volumetric meter at the inlet. When full occlusion occurs the volume displaced starts to decrease, as the roller is rotated away from the measuring device. This is used as indication that the test is complete. The roller volume displacement was measured over a total angle of 50.53~\degree , which is sufficient to cover the entire angle of engagement. For clarification, the angle of engagement is the angle that the pump must rotate for a roller to move from disengaged, but in light contact, to complete occlusion of the tube (or full contact). The test was repeated seven times in order to ensure accuracy and repeatability. 

\section{Experimental design and test bench}
\subsection{Experimental design}
\begin{figure}
\centering
\includegraphics[width = 0.65\textwidth]{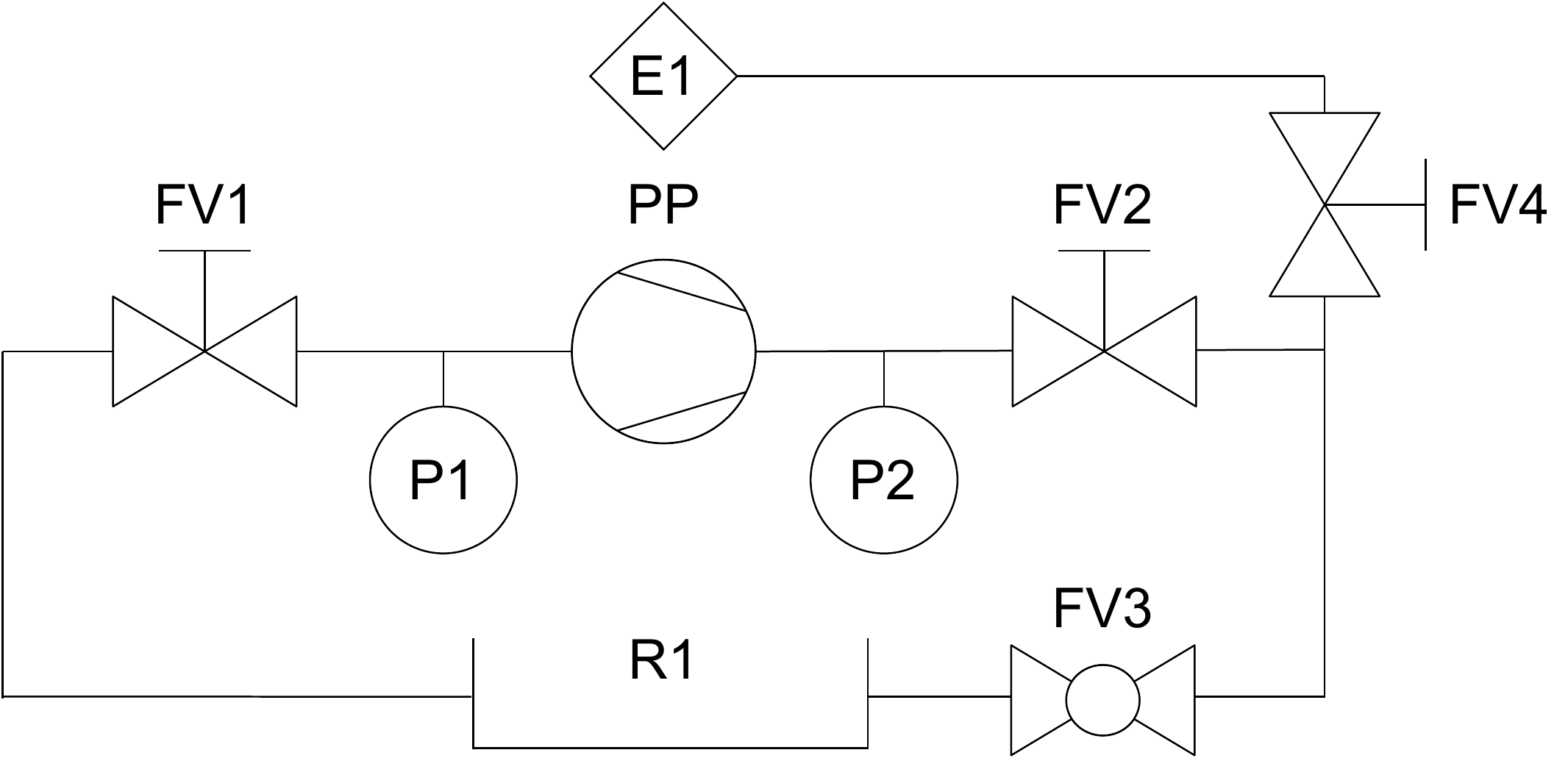}
\caption{P{\&}ID of the test bench for the flow rate tests and pressure tests of the peristaltic pump}\label{PID}
\end{figure}

Fig.~\ref{PID} indicates the piping and instrumentation diagram (P{\&}ID) of the test bench on which two groups of tests are conducted. The first group is to validate the modelled average flow rate of the pump. This group tests the flow rate at various operating speeds of the motor, as well as different operating conditions. The second group of tests are dedicated to the pressure response of the hydraulic circuit. The  method to determine the nominal flow rate of the pump involves pumping water into a measuring beaker (E1) by sealing flow control valve FV3 and opening valve FV4. The second group pumps the water back into the reservoir (R1) via an outlet-line with the same dimensions as the inlet-line by closing valve FV4 and opening valve FV3. 

The flow rate tests comprise five different experiments and are labelled FT\#, with the \# indicating the specific test. The motor is rotated 10 revolutions ($Rev$) at various motor speeds with the mass of the fluid displaced measured using the generic weight scale. The time for the flow rate test using the number of rotations of the pump can be calculated as:
\begin{linenomath*}\begin{equation}
t_{test} = \dfrac{Rev}{N/60}.
\end{equation}\end{linenomath*}

As the flow rate is determined directly from the volume, the accuracy of the modelled volume displacement is equivalent to the accuracy of the flow rate. The process tube is replaced between each test to negate any possible wear effects. The incremental size between motor speeds for each test is 10~r/min. The flow rate test specifications are summarised in the top part of Table~\ref{ft_table}. Some tests are conducted randomly with respect to speed (denoted by R in Table~\ref{ft_table}), while others are conducted systematically with increasing speed (denoted by S in Table 2.

FT1 serves as the control test for the three-roller configuration. FT2 is a secondary control that evaluates the flow rate at intermediate motor speeds of the control. This is done in attempt to achieve a finer resolution of the flow rate at different motor speeds. FT3 tests the flow rate at randomised motor speeds in order to negate the possible occurrence of the process tube heating up during operation from continuous flexing at higher motor speeds. FT4 evaluates the effects of higher operating temperatures of the fluid on the flow rate of the pump. The test examines whether the temperature of the fluid indirectly affects the flow rate by increasing the elasticity of the process tube. These four tests indicate whether there are additional effects taking place inside the pump. FT5 is the control test for the two-roller configuration. In order to maintain a constant water level in the reservoir, the water displaced into the beaker was inserted back into the reservoir after measurement.

The pressure tests comprise two experiments: One for the pressure pulsations for the three-roller pump configuration, the second for the two-roller pump configuration. Both experiments are conducted with the same water level in the reservoir tank (30~cm). Similar to the flow rate tests, the motor speed ranges from 10~r/min to 200~r/min in increments of 10~r/min at room temperature. The pressure tests are, however, not reliant on specific rotations and are therefore tested over a specified time duration (roughly 10~s). The pressure test specifications are summarised in the bottom part of Table~\ref{ft_table} with the three-roller and two-roller pressure tests respectively labelled PT1 and PT2.

\begin{table}[h]
\centering
\caption{Flow rate and pressure test specifications}\label{ft_table}
\begin{tabular}{x{1cm} x{1.2cm} x{1.2cm} x{3cm}}
\toprule
\multirow{2}{*}{\textbf{Label}} & \textbf{No. rollers} & \multirow{2}{*}{\textbf{Temp.}} & \multirow{2}{*}{\textbf{Motor speed}}\\
\midrule
\multicolumn{4}{c}{\textbf{Flow rate tests:}}\\
\midrule
FT1 & 3 & 25~\degree C & 10$\rightarrow$170~r/min (S)\\
\midrule
FT2 & 3 & 25~\degree C  & 15$\rightarrow$165~r/min (S)\\
\midrule
FT3 & 3 & 25~\degree C  & 10$\rightarrow$170~r/min (R)\\
\midrule
FT4 & 3 & 40~\degree C  & 10$\rightarrow$170~r/min (R)\\
\midrule
FT5 & 2 & 25~\degree C  & 10$\rightarrow$300~r/min (S)\\
\midrule
\multicolumn{4}{c}{\textbf{Pressure tests:}}\\
\midrule
PT1 & 3 & 25~\degree C & 10$\rightarrow$200~r/min (S)\\
\midrule
PT2 & 2 & 25~\degree C & 10$\rightarrow$200~r/min (S)\\
\bottomrule
\end{tabular}
\end{table}

\subsection{Test bench}
\begin{figure}[t]
\centering
\includegraphics[width = .9\textwidth]{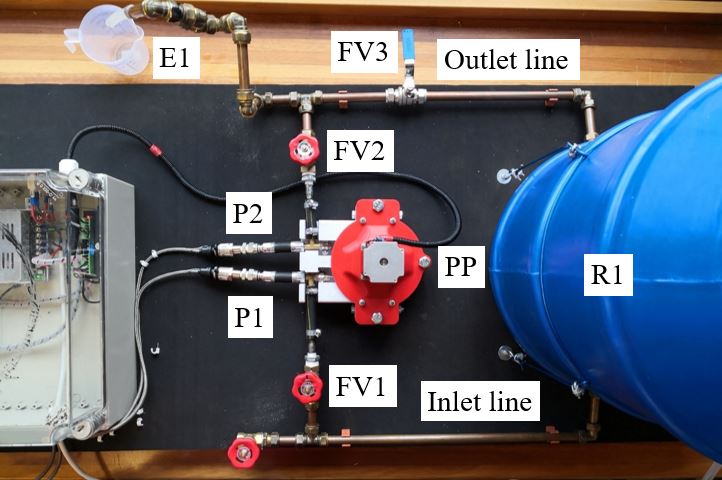}
\caption{Peristaltic pump test bench developed by the North-West University, the following symbols indicate: FV - flow control valve, P - Pressure sensor, PP - peristaltic pump, E1 - measuring beaker, R1 - reservoir}\label{TB}
\end{figure}

In order to validate the revised model, a validation test bench was developed to work in conjunction with the developed roller-type peristaltic pump as indicated in Fig.~\ref{TB}. The test bench sensors consist of two Honeywell PX2EF1XX050P-AAAX pressure sensors and a K-type thermocouple. The pressure sensors are located directly at the inlet ($P_{in}$) and outlet ($P_{out}$) of the peristaltic pump. A MAX 6675 breakout board is used to gather information from the thermocouple and transmit it to the data acquisition unit.

The data acquisition and control was accomplished by means of a desktop computer with a dSPACE 1104 R \& D controller board (dS 1104) connected via PCIe. The dSPACE control and acquisition was done in the ControlDesk virtual environment. Additionally, the North-West University developed an open-source general purpose interface board for the dPSACE 1104 board, available at \url{www.eetips.xyz}. 

In order to measure the volume flow rate in a cost effective and accurate manner, the outlet of the pump was connected to a measuring beaker. The fluid displaced into the beaker after a test was weighed after the pump ran at a certain speed for a designated amount of rotations, and thus time. The stepper motor was controlled via a Wantai DQ542MA hybrid stepper motor driver controlled via the digital-to-frequency (D2F) port of the dSPACE board. The pressure sensors were calibrated using a testo 622 hygrometer and barometer.

The flow control valves FV1 and FV2 are used to completely seal off the inlet or outlet in order to test the maximum and minimum working pressure of the pump. The pressure sensor P1 is commonly associated with the inlet-line of the pump (PP). However, as the pump is unidirectional, it is subject to change based on the direction. The same yields true for the pressure sensor P2 and the outlet-line. The flow control valve FV3 is used during the flow rate tests in order to isolate the peristaltic pump outlet.

The pressure sensors have a response time of 2~ms, however, the pressure validation tests had a sample time of 0.2~ms (or a sample rate of 5~kHz). The sample rate and response time of the pressure sensors are sufficient, as the expected time for the motor to rotate an angle of 50.53~\degree ~(the span of the angle of engagement) at 200~r/min is 470.4~ms.

The fluid resistance of the inlet-line (and thus also the outlet-line) was determined using a kinetic pump, pressure sensors, and a WeiHeng weight scale. The pressure sensors are located at the inlet and outlet of the test bench's isolated inlet-line. The kinetic pump, which was used for its steady flow rate, was operational for one minute, using a digital timer and a solenoid switch. The timer was set to a minute to observe the average flow rate over a larger span of time, mitigating the effects of ramp-up flow rates during initialisation. The longer testing time also mitigates weight sensor inaccuracy for lower masses. An ATmega2560 was used for timing, operation, and data acquisition for both the resistance and compliance tests due to operational restrictions. The volume displaced by the pump during operation gives an indication of the flow rate of the pump through the inlet-line. The resistance could then be determined with the pressure differential between the two pressure sensors and flow rate as~\cite{karnopp2012system}:

\begin{linenomath*}\begin{equation}
R  = \dfrac{\Delta P}{Q}.
\end{equation}\end{linenomath*}

The flow rate can be determined by means of the measured mass displacement ($m$) using the WeiHeng scale as:
\begin{linenomath*}\begin{equation}
\label{Q-t}
Q_{test} =  \dfrac{V_{test}}{t_{test}},
\end{equation}\end{linenomath*}
with:
\begin{linenomath*}\begin{equation}
\label{V-t}
V_{test} = \dfrac{m}{\rho},
\end{equation}\end{linenomath*}
with $\rho$ denoting the density of the water (997~kg/m$^3$ at 25~\degree C and 992~kg/m$^3$ at 40~\degree C \cite{Menon2015}).  The compliance of the test bench's inlet-line is tested by sealing the inlet and outlet of the isolated inlet-line, filled with water, and injecting additional water with a syringe. The rise in pressure due to the change in volume is recorded before and after the additional volume injection. The change in pressure with the change in volume is indicative of the compliance of the tube. The effective bulk modulus ($B_e$) of the water and the pipe is determined as~\cite{karnopp2012system}:

\begin{linenomath*}\begin{equation}
B_e = \dfrac{\Delta P\cdot V_{total}}{\Delta V},
\end{equation}\end{linenomath*}

\noindent with $\Delta V$ the volume that was injected, and $V_{total}$ the total volume in the isolated inlet-line and syringe. The effective compliance of the inlet and outlet-lines ($C$) can then be determined with~\cite{karnopp2012system}:

\begin{linenomath*}\begin{equation}
C = \dfrac{1}{B_e}.
\end{equation}\end{linenomath*}

The inertia of the hydraulic circuit is iteratively solved for the 100~r/min experimental test using the simulation parameters. The inertia value yielding the closest simulation values to that of the experimental results is used for all simulations.

\section{Results and discussion}
The simulation of the pressure response was conducted with the parameters indicated in Table~\ref{simulation_table_param} for both the two-roller and three roller configurations. The modelled flow rate ($Q_{avg}$) is then compared to the experimental flow rate for both the two-roller and three-roller configurations for each motor speed value. The model and simulation values are compared to the practical values using the root mean square error~(RMSE), normalised root mean square error~(NRMSE), and the Pearson correlation coefficient~($\rho_{\text{X,Y}}$) where applicable.

\begin{table}[ht!]
\centering
\caption{Simulation parameters used to validate the lumped parameter model}\label{simulation_table_param}
\begin{tabular}{p{2.9cm} p{0.1cm}p{.7cm} x{1.15cm} x{1.7cm}}
\toprule
\textbf{Parameter:} & \multicolumn{2}{c}{\centering \textbf{Symbol:}} & \textbf{Value:}& \textbf{Unit:}\\
\midrule
Reservoir pressure & &$P_{res}$ & 88.637 & kPa\\
Inlet resistance & &$R_{in}$ 	& 0.1108 & kPa$\cdot$s/mL\\
Outlet resistance & &$R_{out}$  & 0.1108 & kPa$\cdot$s/mL\\
Inlet capacitance &&$C_{in}$ 	& 0.0361 & mL/kPa\\
Outlet capacitance &&$C_{out}$  & 0.0361 & mL/kPa\\
Inlet inductance && $L_{in}$ 	& 0.0042 & kPa$\cdot$s$^2$/mL\\
Outlet inductance & &$L_{out}$  & 0.0042 & kPa$\cdot$s$^2$/mL\\
\bottomrule
\end{tabular}
\end{table}

\subsection{Flow rate}
Ten volume samples were taken at each motor speed in order to ensure repeatability and accuracy. The standard deviation~(STD) for the room temperature tests with motor speeds below 160~r/min are all below 1~mL. The probability of measuring larger volumes that what the pump displaces is far less than measuring less volume with regards to testing conditions. For this reason the maximum values of the ten samples are used for evaluation and comparison. The effects of the elevated temperature does not show a significant effect with regards to the volume displaced.

\begin{figure}[t!]
\centering
\begin{tabular}{c}
\subfloat[Three-roller volume displacement]{\includegraphics[width = 0.8\textwidth ,trim={3.4cm 10.3cm 4cm 10.9cm},clip]{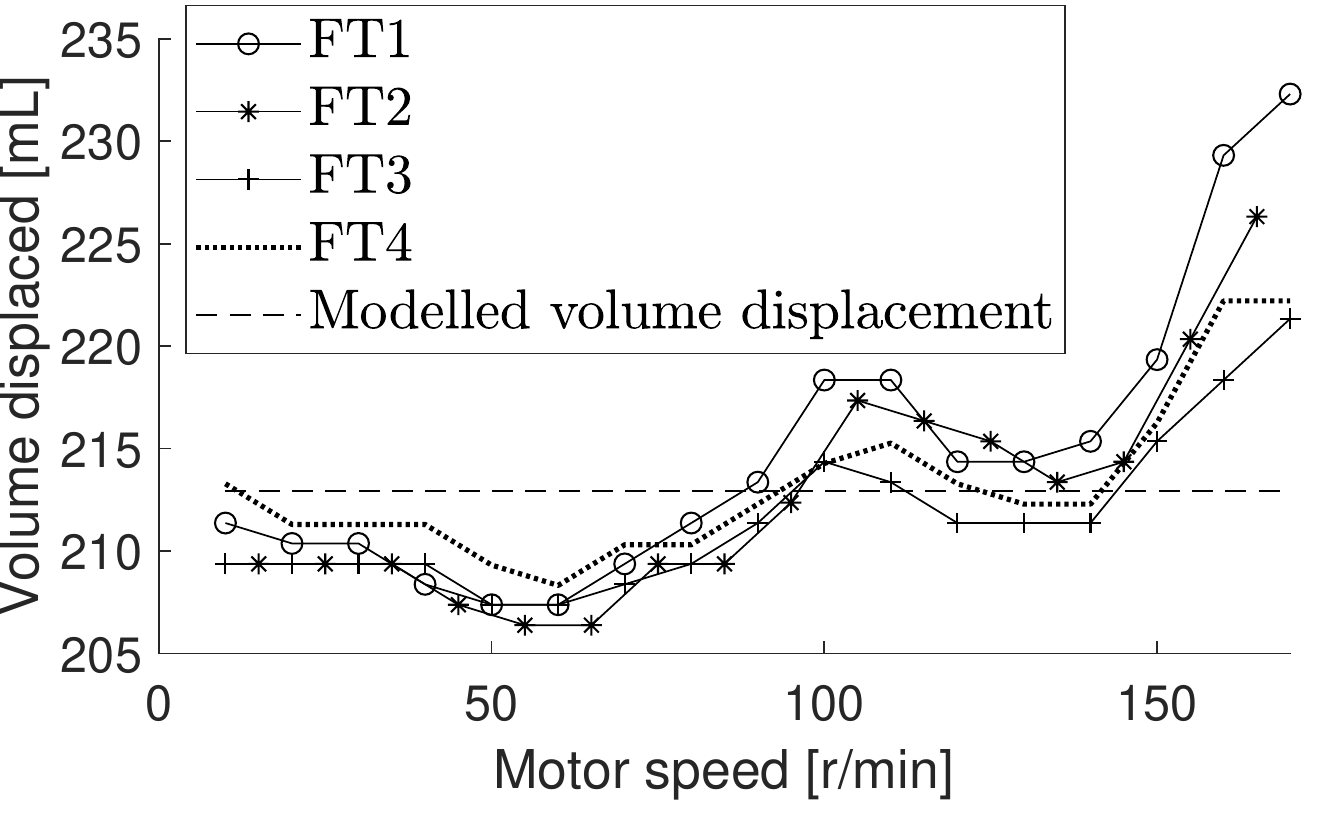}\label{comp_vol1}}\\
\subfloat[Two-roller volume displacement]{\includegraphics[width = 0.8\textwidth ,trim={3.4cm 11.2cm 4cm 11.5cm},clip]{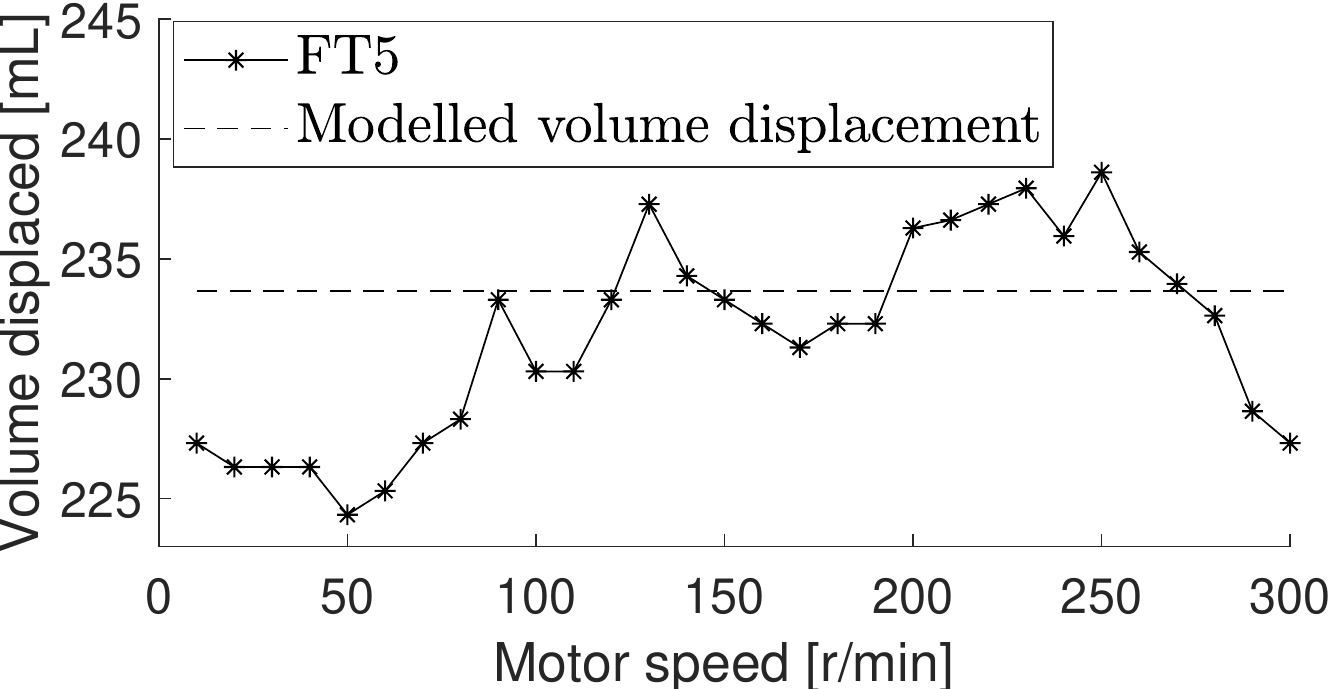}\label{comp_vol2}}
\end{tabular}
\caption{Pump volume displacement test results for 10 rotations at various motor speeds compared to the modelled values for the (a) three-roller configuration and (b) two-roller configuration}\label{comp_vol}
\end{figure}

\begin{figure}[!h]
\centering
\begin{tabular}{c}
\subfloat[Three-roller configuration]{\includegraphics[width = 0.8\textwidth  ,trim = {3.3cm 11.55cm 3.5cm 11.6cm},clip]{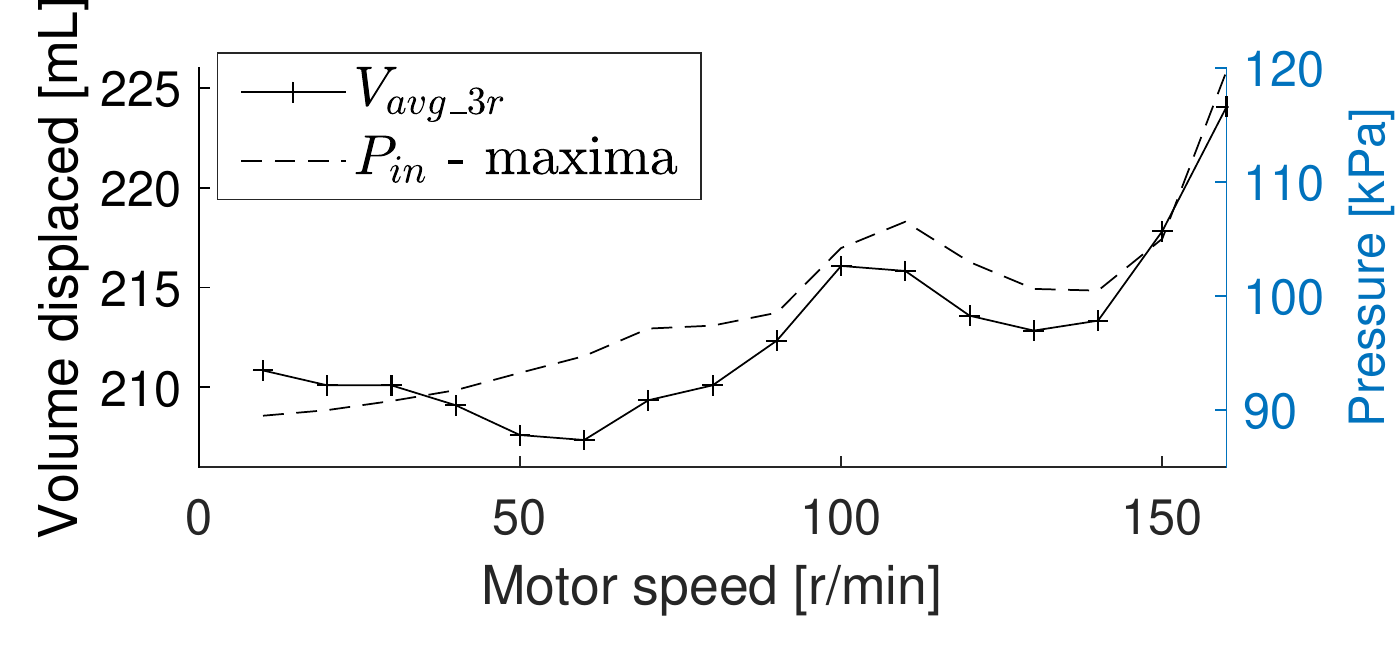}\label{peak_cor}}\\
\subfloat[Two-roller configuration]{\includegraphics[width = 0.8\textwidth ,trim = {3.3cm 11.55cm 3.5cm 11.6cm},clip]{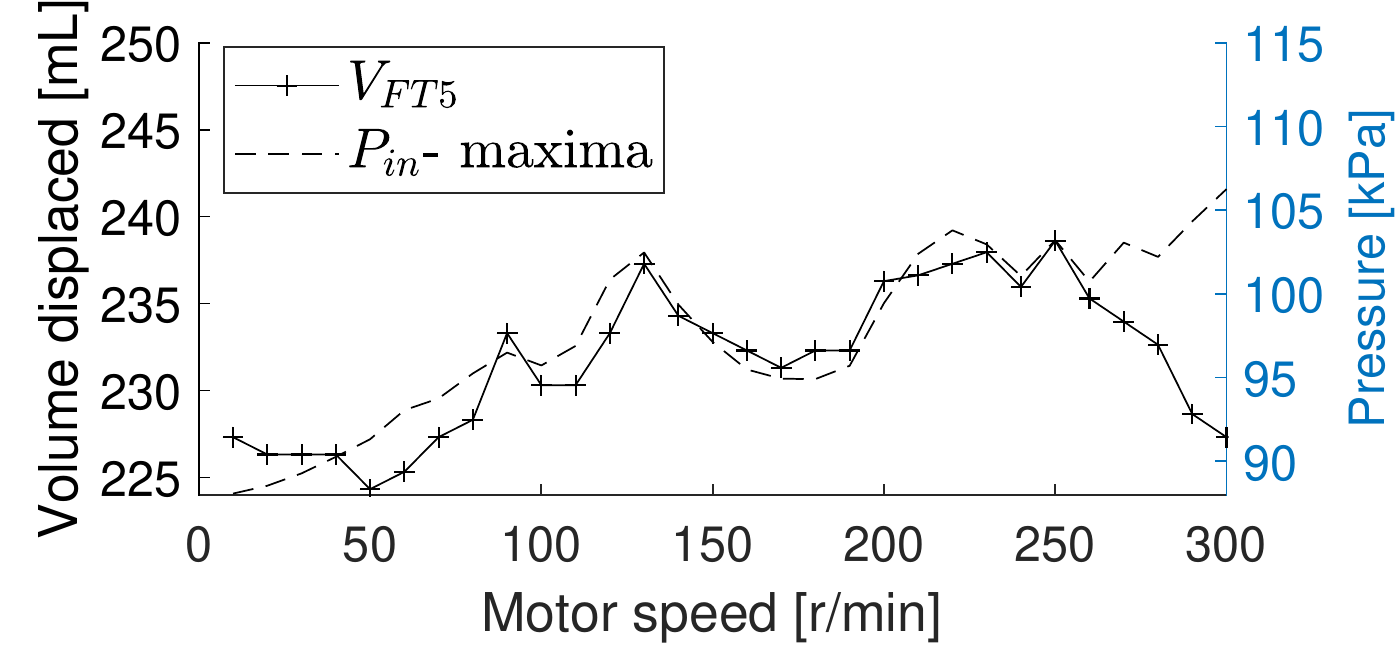}\label{v_vsP_d}}
\end{tabular}
\caption{Volume displacement results for the (a) three-roller and (b) two-roller volume tests compared to the inlet peak maximum pressures}
\label{peak_cor_true}
\end{figure}

The modelled volume displacement of the pump for the three-roller configuration, indicated by the dashed line in Fig.~\ref{comp_vol1}, had a normalised root mean square error~(NRMSE) of 2.37~\%. This value is determined for the average of the three-roller configuration tests ($V_{avg\_3r}$). The NRMSE for the modelled volume displacement for the two-roller configuration, indicated by the dashed line in Fig.~\ref{comp_vol2}, was slightly lower at 1.97~\%. Additionally, the volume displacement is not linear with respect to the motor speed of the pump. The volume displacement of both the two-roller and three-roller configurations correlate strongly to the peak inlet pressure of the pump, indicated in Fig.~\ref{peak_cor_true}. The modelled volume displacement of both configurations show considerable accuracy, even with the non-linear effects not accounted for. The largest deviations from the experimental values are roughly 9.02~\% for the three-roller configuration and 4.03~\% for the two-roller configuration. 

The accuracy of the modelled volume displacement for the three-roller configuration is better for lower motor speeds than that of the two-roller configuration. The two-roller configuration is also noted to deliver larger volumes, with an average of 231.75~mL, than that of the three-roller configuration, with an average of 212.53~mL, as expected. 

\subsection{Pressure pulsations}
\begin{figure}[!h]
\hspace*{-0.3cm}
\begin{tabular}{cc}
\subfloat[]{\includegraphics[width = 0.4\textwidth , trim={3.7cm 9.35cm 3.5cm 9.8cm},clip]{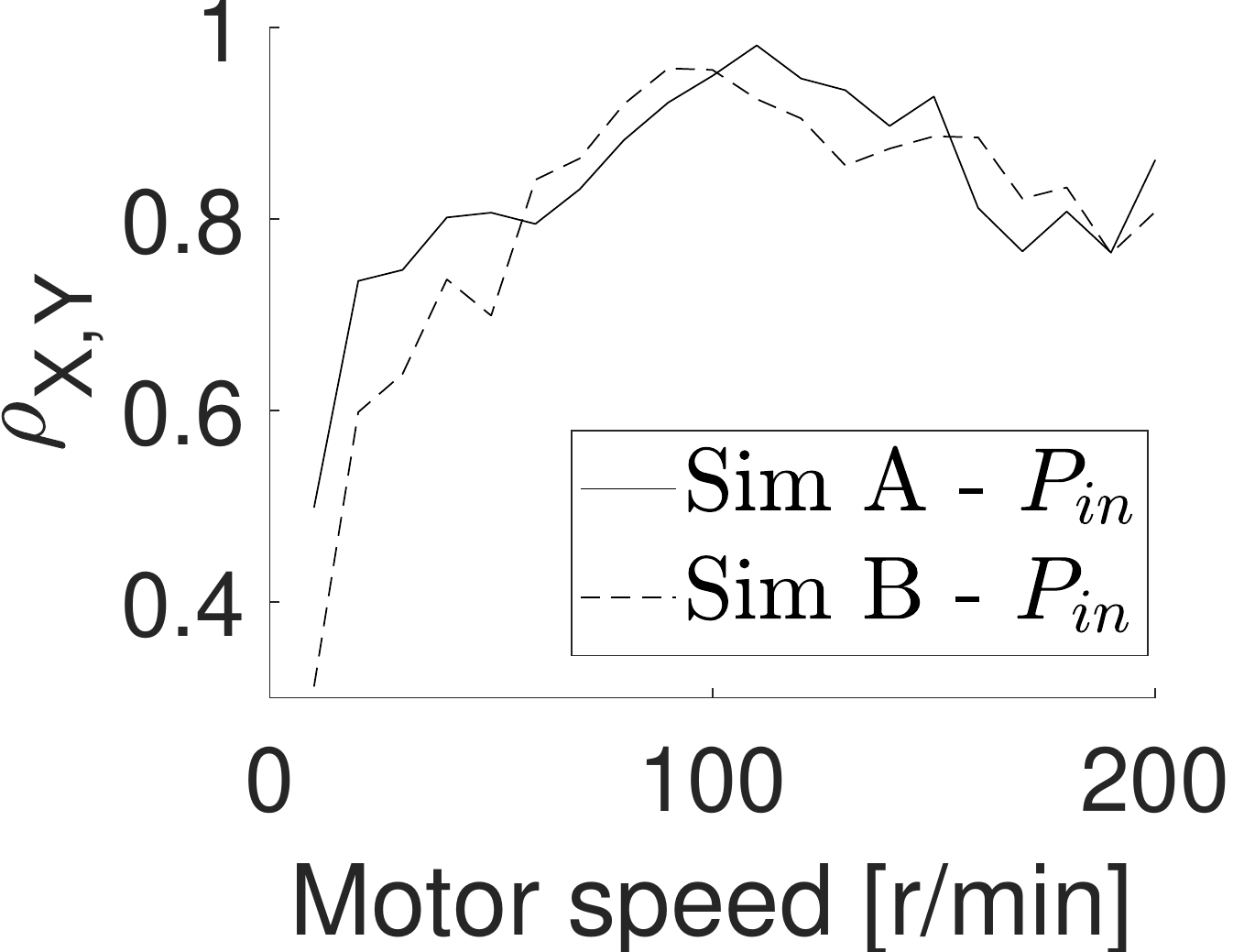}}
\subfloat[]{\includegraphics[width = .4\textwidth , trim={3.7cm 9.35cm 3.5cm 9.8cm},clip]{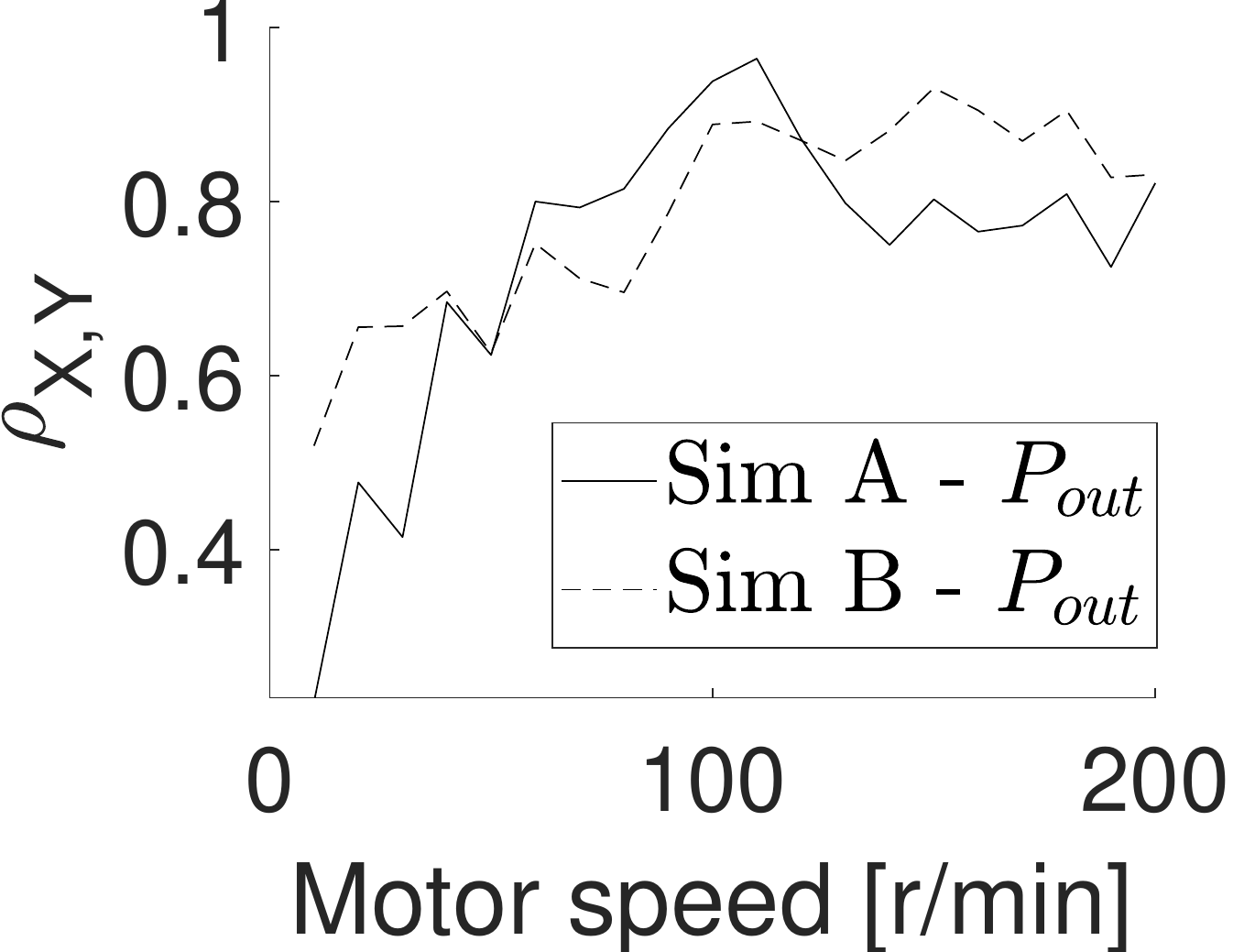}}
\end{tabular}
\caption{Pearson correlation coefficient of the simulated (a)~inlet and (b)~outlet pressures over various motor speeds for the three-roller (Sim A) and two-roller (Sim B) configurations}\label{cor_coef}
\end{figure}

\begin{figure*}[!h]
\centering
\begin{tabular}{ccc}
\subfloat[Inlet pressure ($P_{in}$) at 50~r/min]{\includegraphics[width = 0.32\textwidth ,trim={3.3cm 9.3cm 4cm 9.8cm},clip]{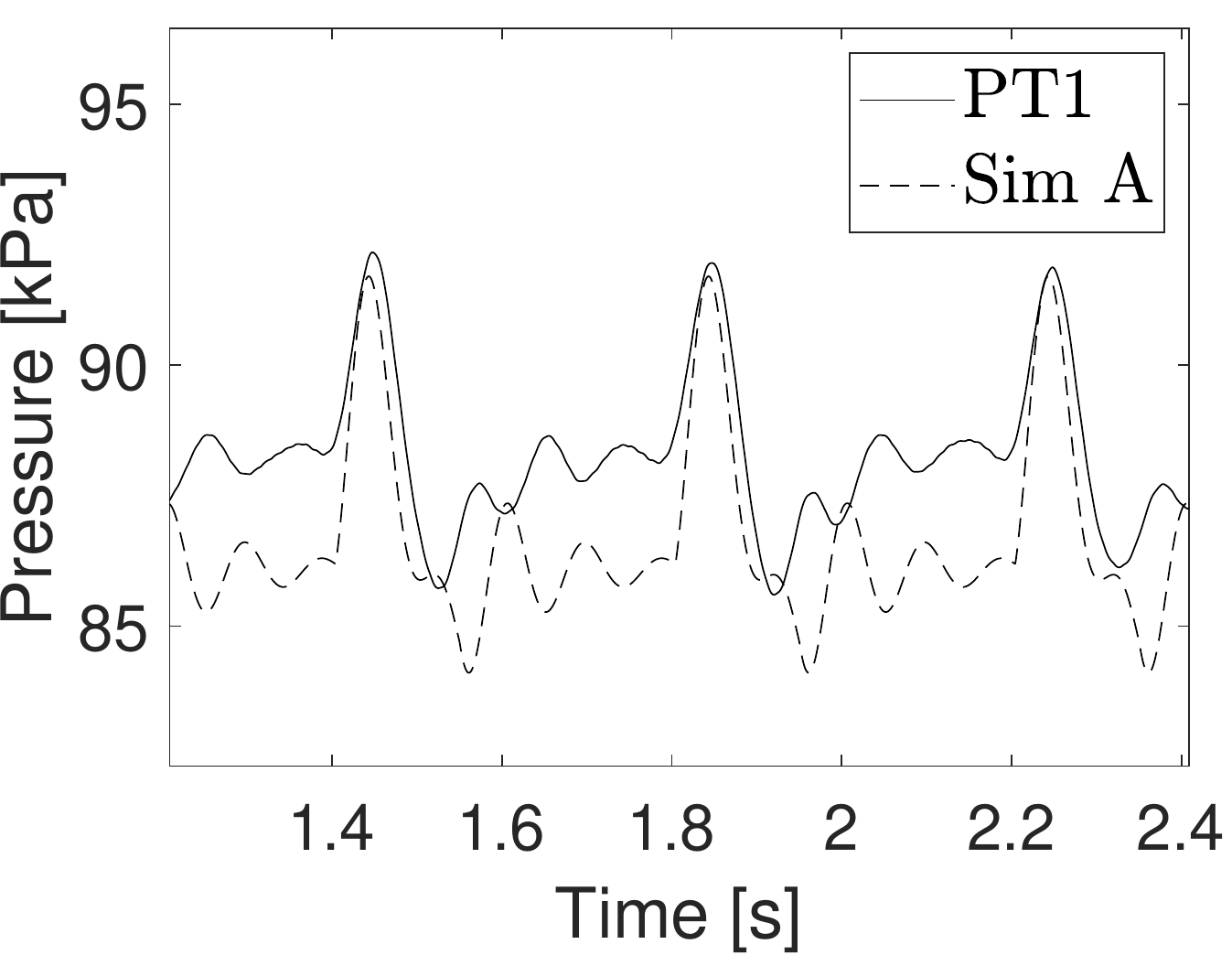}\label{3r_50_p1}} &
\subfloat[Inlet pressure ($P_{in}$) at 100~r/min]{\includegraphics[width = 0.32\textwidth ,trim={3.3cm 9.3cm 4cm 9.8cm},clip]{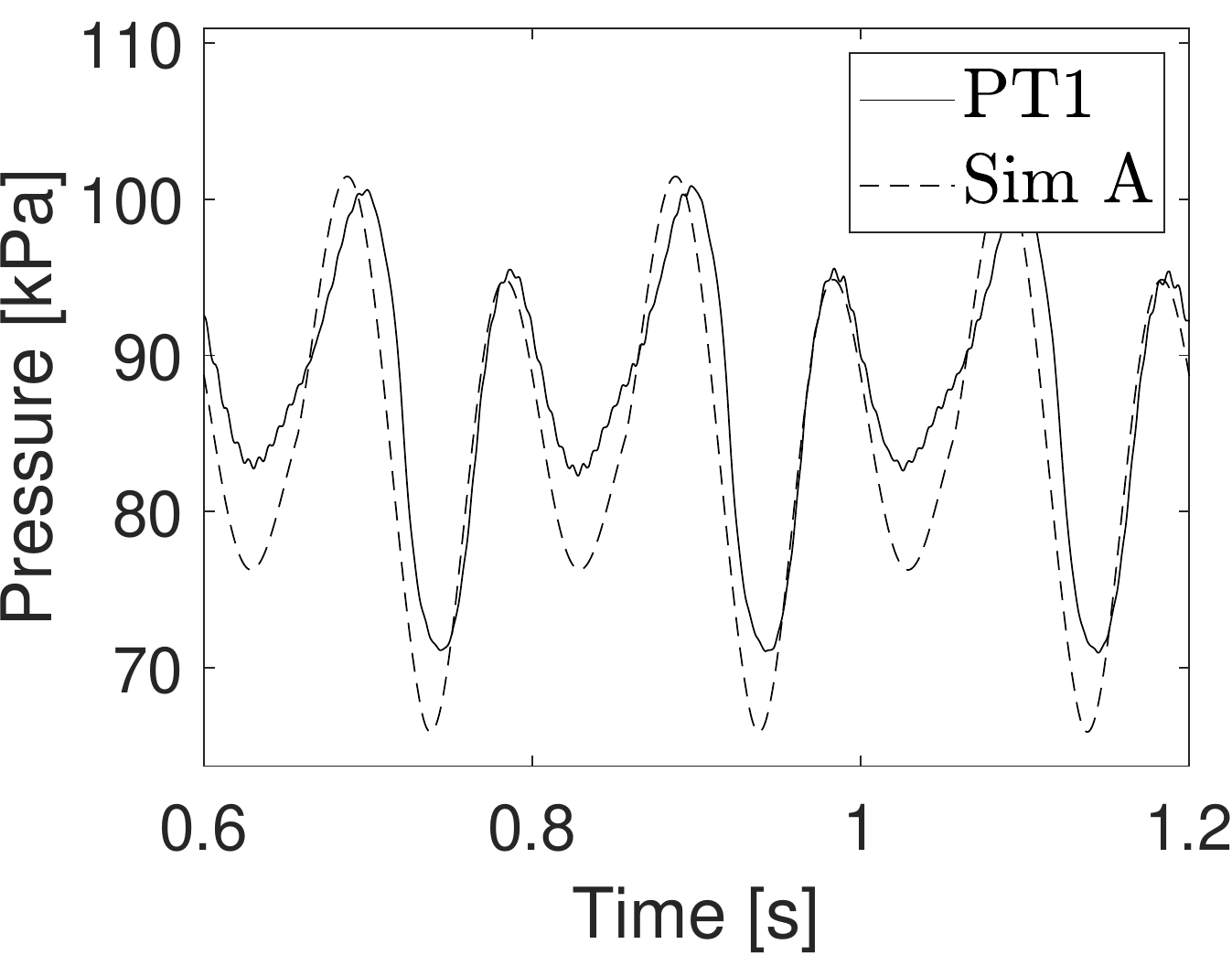}} &
\subfloat[Inlet pressure ($P_{in}$) at 150~r/min]{\includegraphics[width = 0.32\textwidth ,trim={3.3cm 9.3cm 4cm 9.8cm},clip]{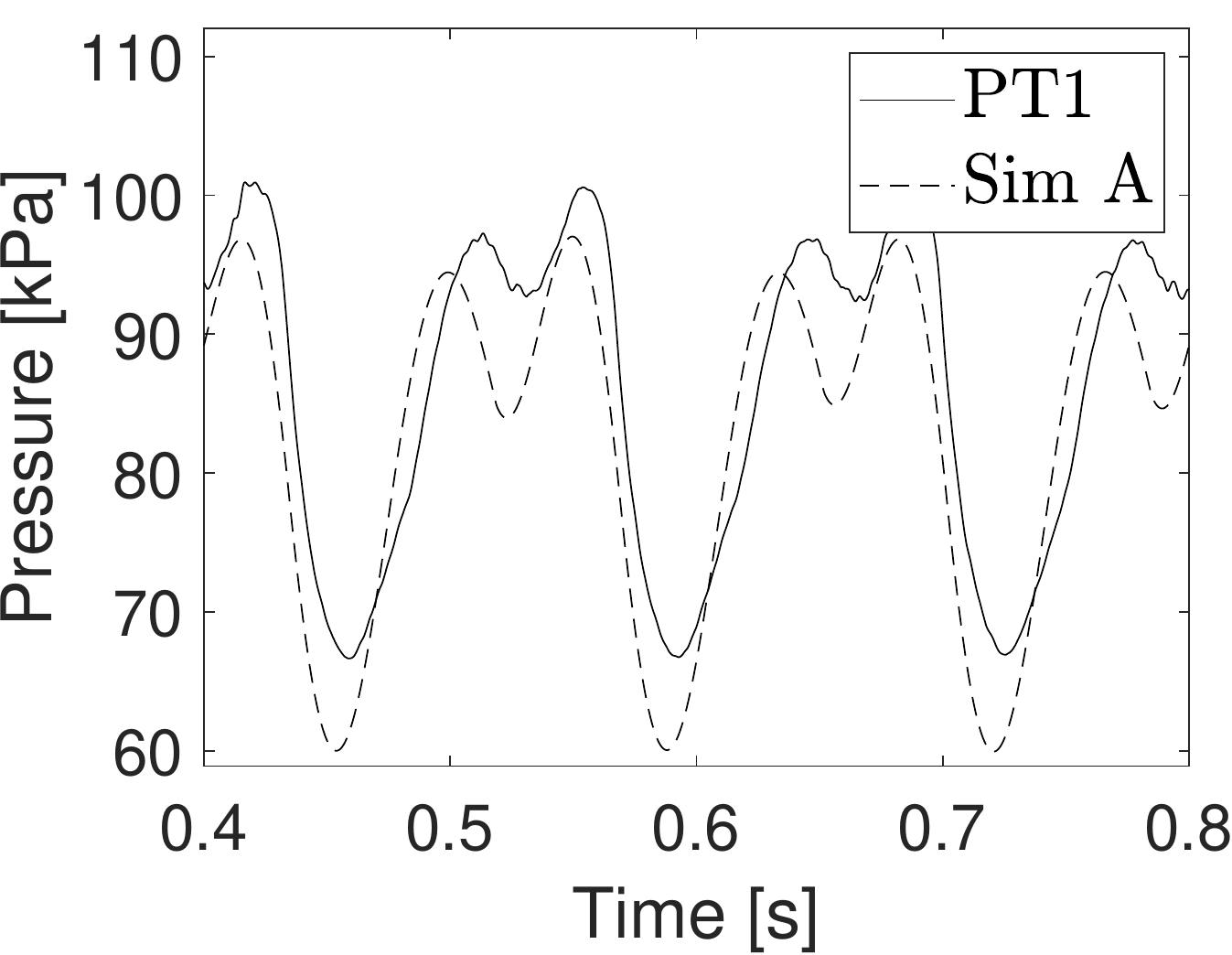}\label{3r_150_p1}}\\
\subfloat[Outlet pressure ($P_{out}$) at 50~r/min]{\includegraphics[width = 0.32\textwidth,trim={3.3cm 9.3cm 4cm 9.8cm},clip]{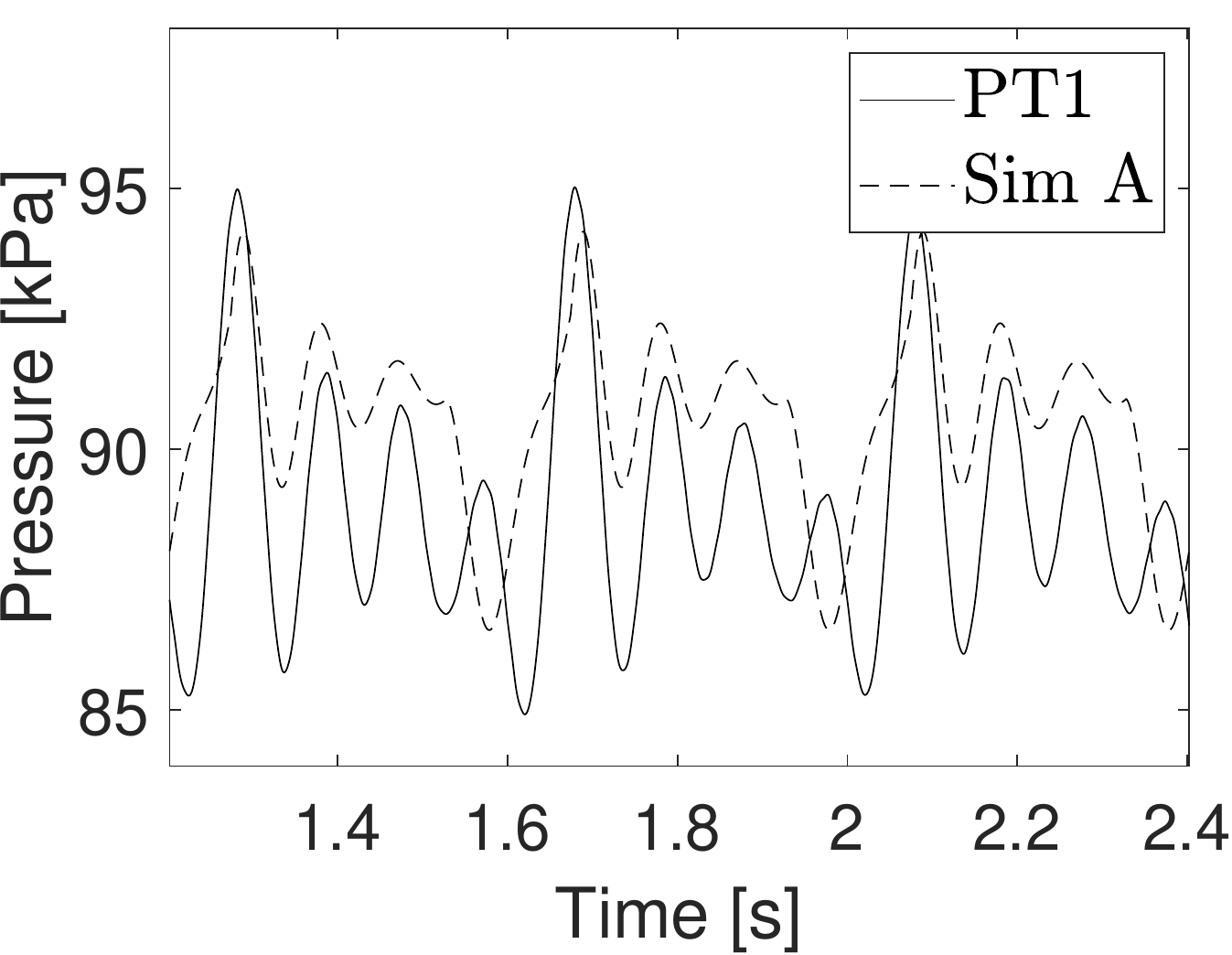}\label{3r_50_p2}}&
\subfloat[Outlet pressure ($P_{out}$) at 100~r/min]{\includegraphics[width = 0.32\textwidth,trim={3.3cm 9.3cm 4cm 9.8cm},clip]{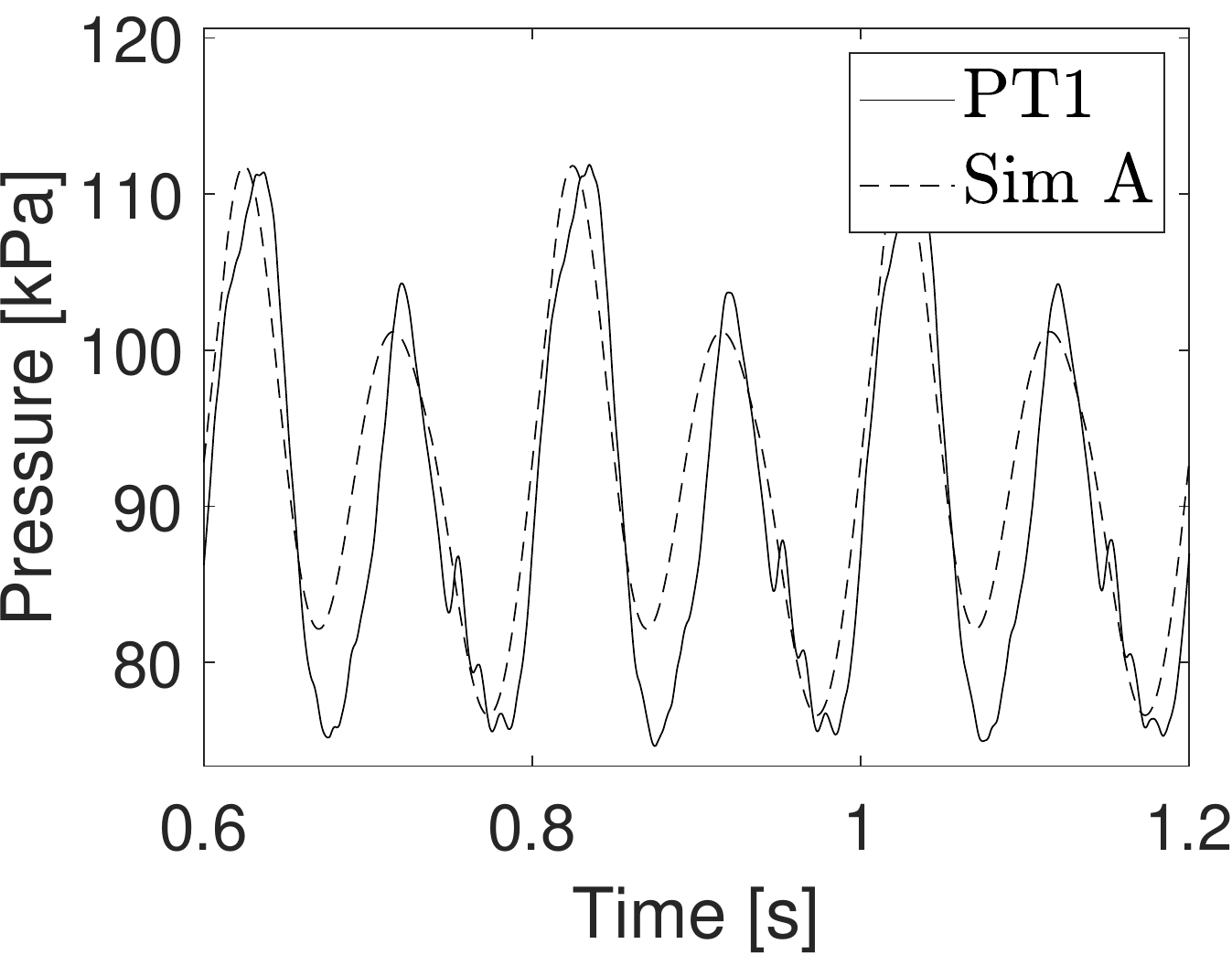}}&
\subfloat[Outlet pressure ($P_{out}$) at 150~r/min]{\includegraphics[width = 0.32\textwidth,trim={3.3cm 9.3cm 4cm 9.8cm},clip]{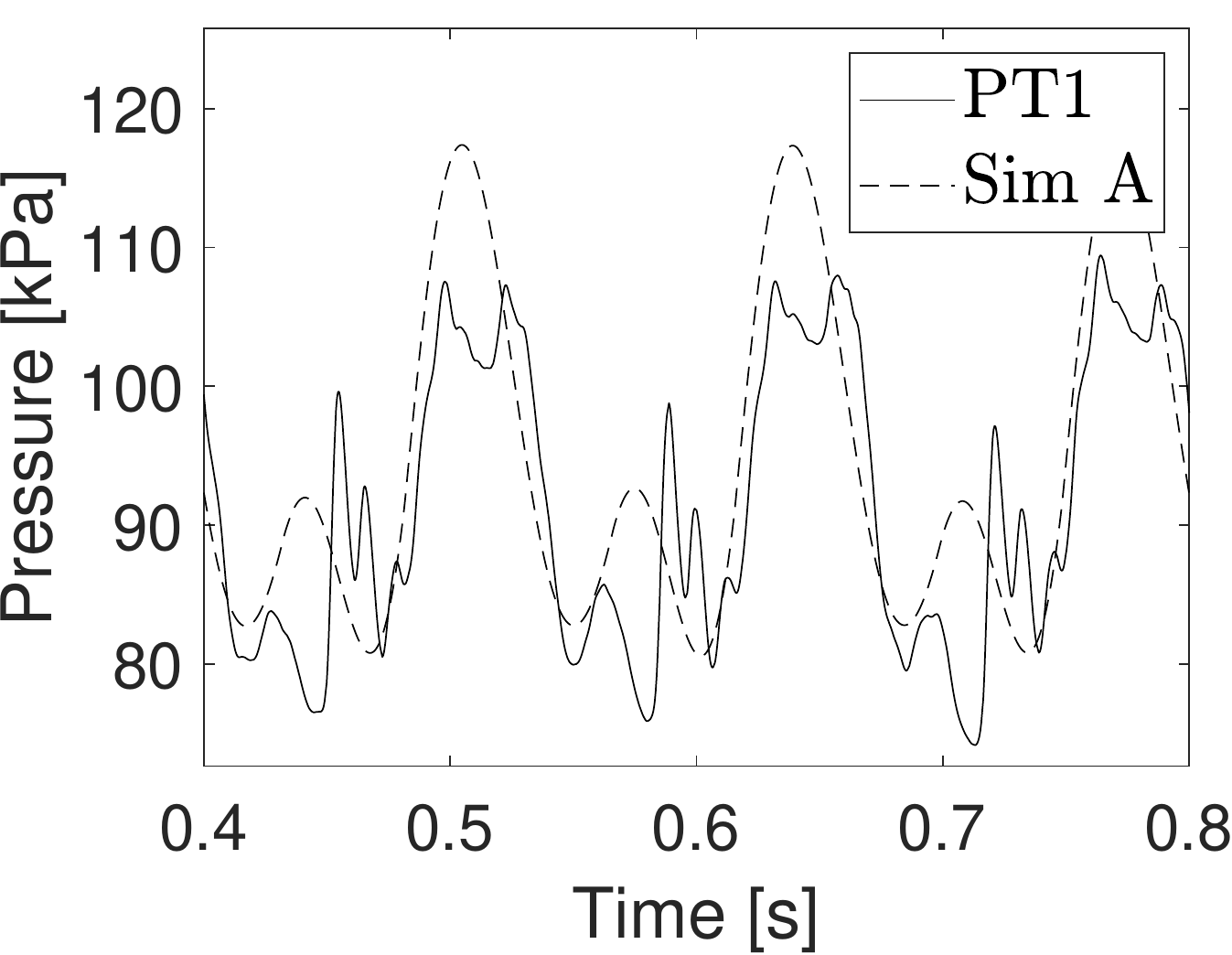}\label{3r_150_p2}}\\
\end{tabular}
\caption{Three-roller experimental (PT1) and simulation (Sim A) results for various motor speeds}\label{sim_correl}
\end{figure*}

\begin{figure*}[!h]
\centering
\begin{tabular}{ccc}
\subfloat[Inlet pressure ($P_{in}$) at 50~r/min]{\includegraphics[width = 0.32\textwidth ,trim={3.3cm 9.3cm 4cm 9.8cm},clip]{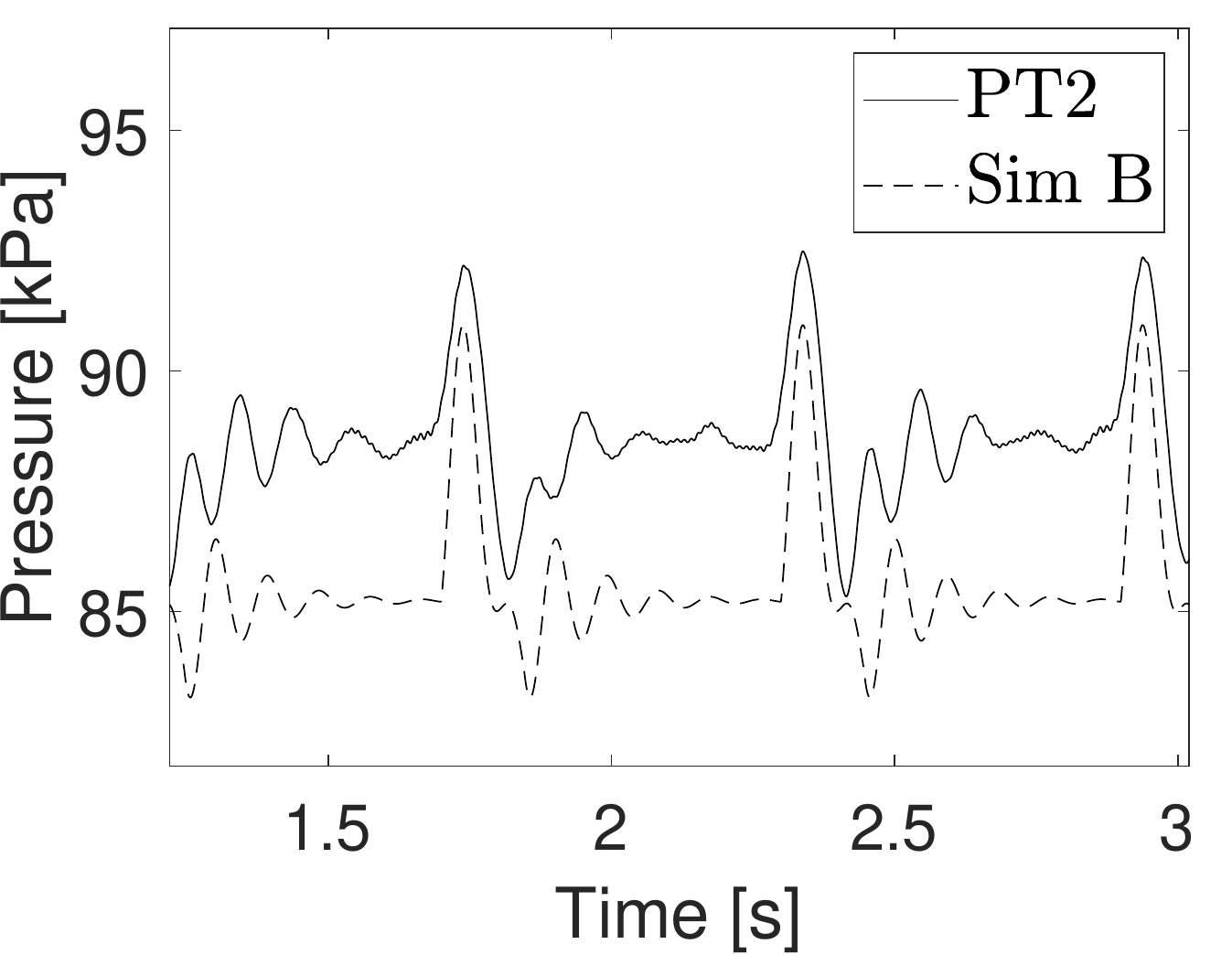}\label{2r_50_p1}} &
\subfloat[Inlet pressure ($P_{in}$) at 100~r/min]{\includegraphics[width = 0.32\textwidth ,trim={3.3cm 9.3cm 4cm 9.8cm},clip]{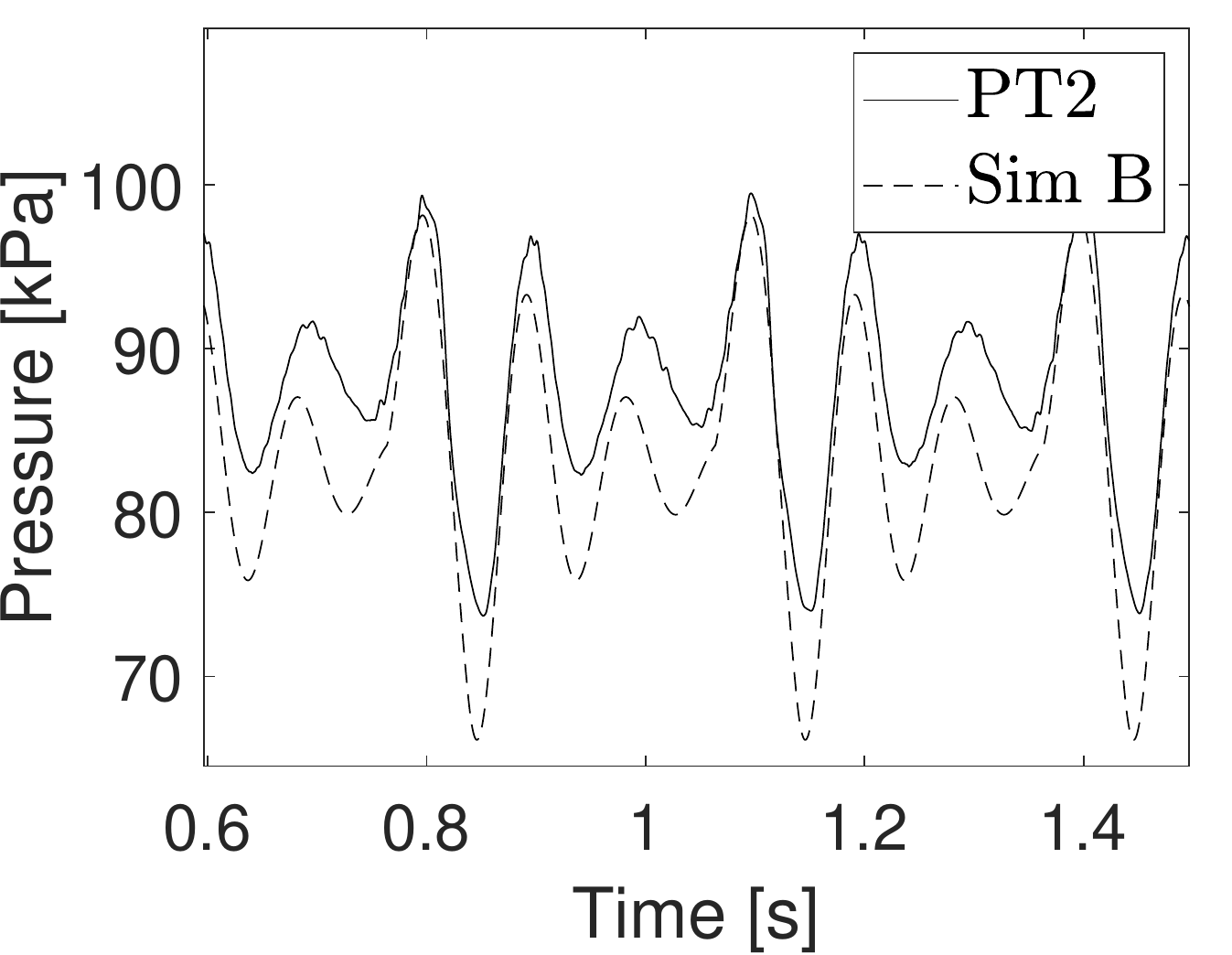}} &
\subfloat[Inlet pressure ($P_{in}$) at 150~r/min]{\includegraphics[width = 0.32\textwidth ,trim={3.3cm 9.3cm 4cm 9.8cm},clip]{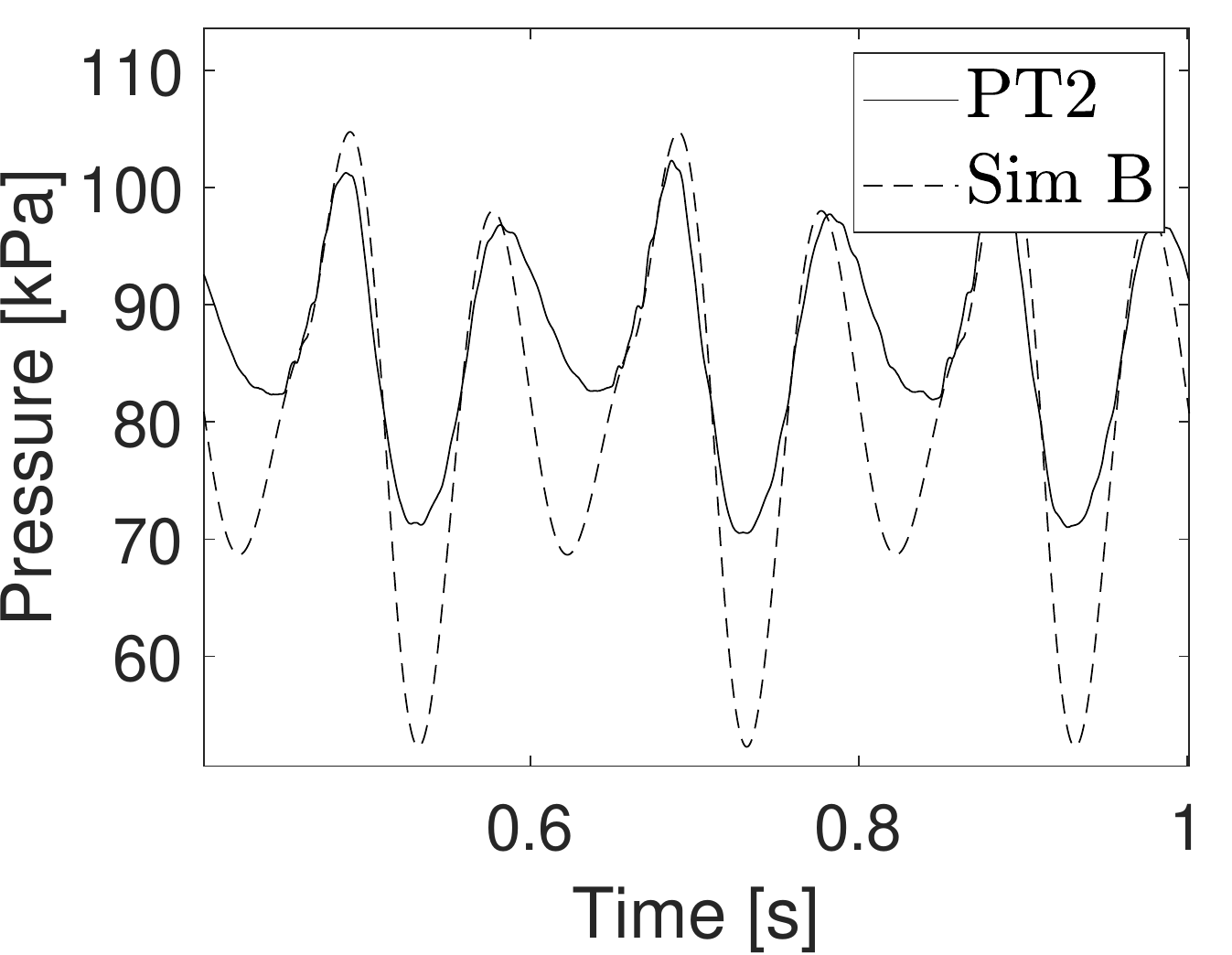}\label{2r_150_p1}}\\
\subfloat[Outlet pressure ($P_{out}$) at 50~r/min]{\includegraphics[width = 0.32\textwidth,trim={3.3cm 9.3cm 4cm 9.8cm},clip]{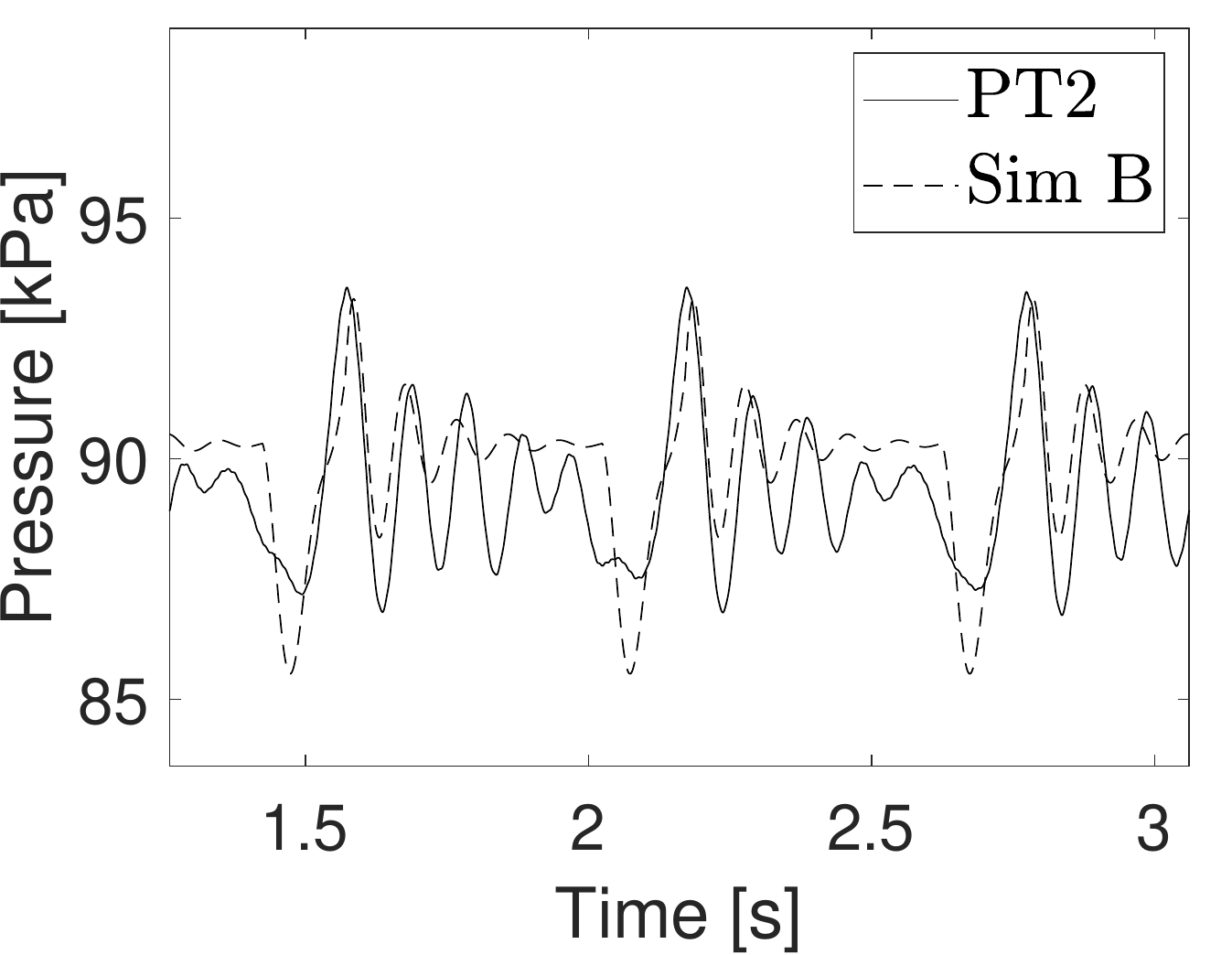}\label{2r_50_p2}}&
\subfloat[Outlet pressure ($P_{out}$) at 100~r/min]{\includegraphics[width = 0.32\textwidth,trim={3.3cm 9.3cm 4cm 9.8cm},clip]{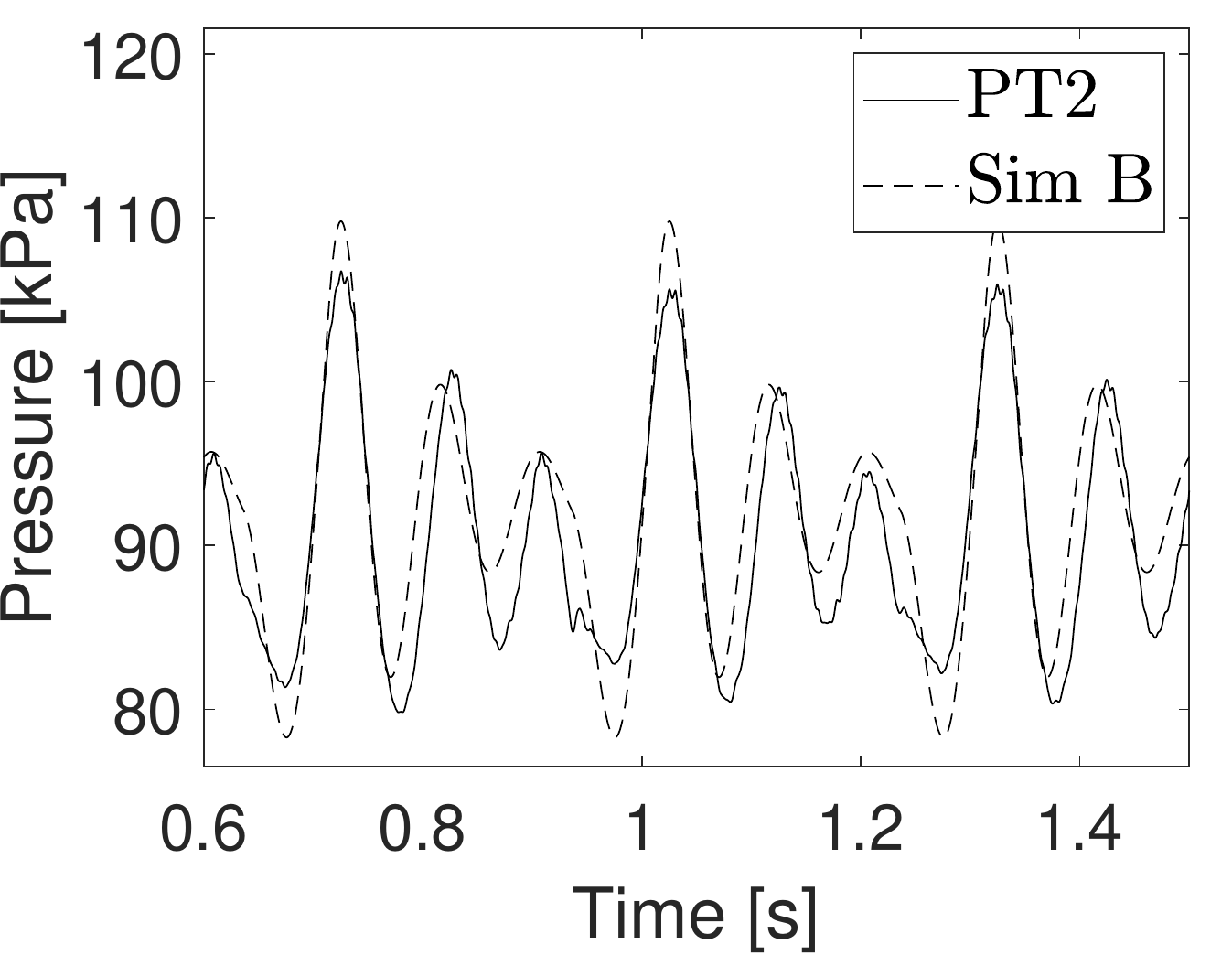}}&
\subfloat[Outlet pressure ($P_{out}$) at 150~r/min]{\includegraphics[width = 0.32\textwidth,trim={3.3cm 9.3cm 4cm 9.8cm},clip]{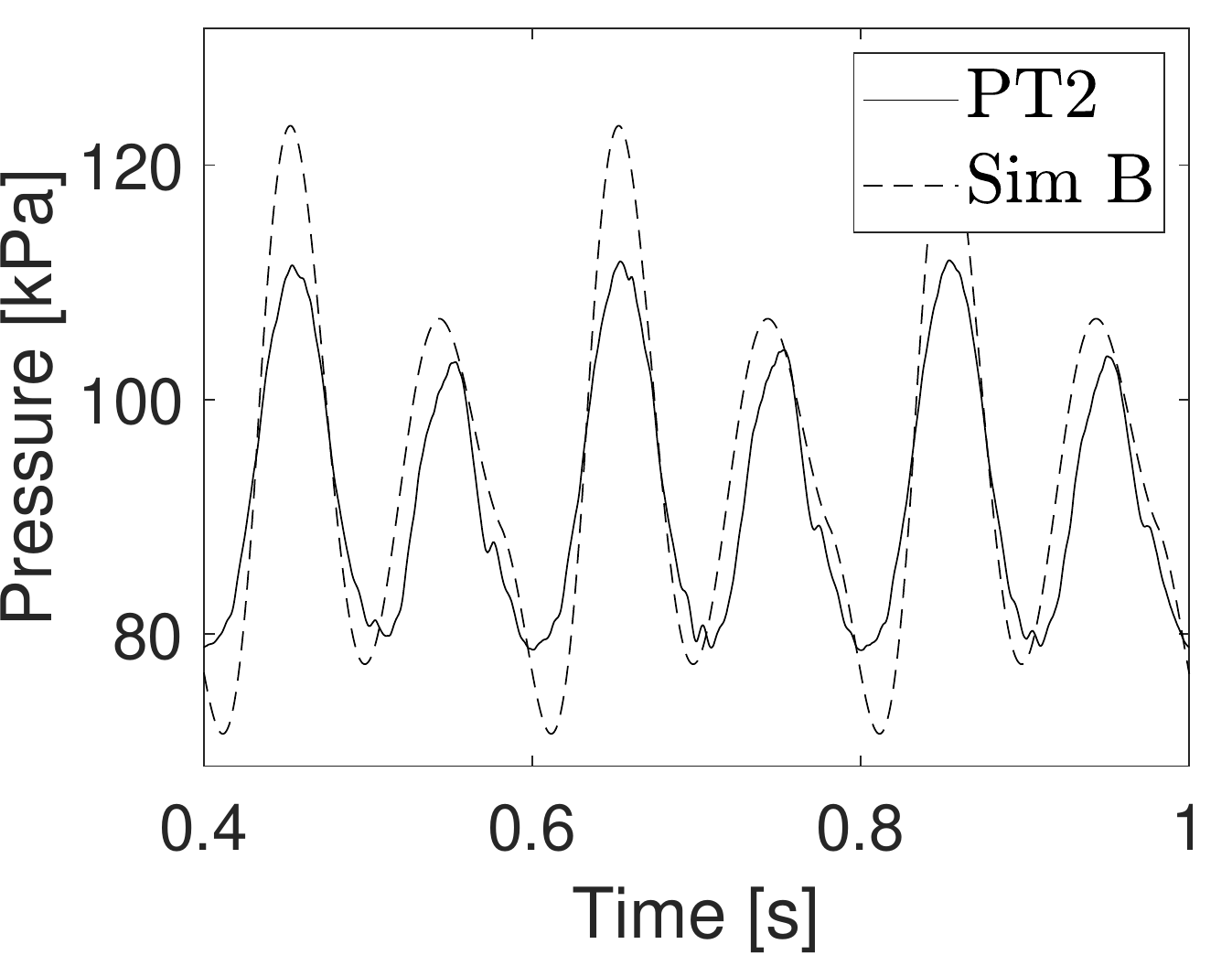}\label{2r_150_p2}}\\
\end{tabular}
\caption{Two-roller experimental (PT2) and simulation (Sim B) results for various motor speeds}\label{sim_correl2}
\end{figure*}

Both the simulations and experimental results of PT1 and PT2 are shown in Figs.~\ref{sim_correl} and~\ref{sim_correl2} respectively for motor speeds of 50, 100, and 150~r/min. The Pearson correlation coefficient ($\rho_{_{\text{X,Y}}}$) is calculated for the full duration of the tests for each motor speed and indicated in Fig.~\ref{cor_coef}. The correlation coefficient provides information on how well the simulation waveform matches that of the experiment. The correlation coefficient indicates an exact match with a value of 1, no correlation with a value of 0, and negative correlation with a value of -1. Correlation coefficient values above 0.7 are considered strong correlations~\cite{schober2018correlation}. The simulation pertaining to the three-roller configuration is denoted as Sim A and that of the two-roller configuration as Sim B.

The simulation is noticeably less accurate at lower motor speeds (10-40~r/min), however, shows strong correlation ($\rho_{_\text{X,Y}}>0.7$) above 50~r/min for both configurations. This may be due to the low signal-to-noise ratio at lower motor speeds, or due to the fixed parameter values of the simulated lumped parameter model. The fixed values of the inlet and outlet-line's resistance were experimentally determined at larger flow rates (roughly 210~mL/s). Furthermore, the inertia remained fixed for all simulations with the value determined from the closest fit for the 100~r/min test. This can result in larger discrepancies as offsets at lower flow rates as seen in the 50~r/min tests in Figs.~\ref{3r_50_p1}, \ref{3r_50_p2}, \ref{2r_50_p1}, and~\ref{2r_50_p2}. Additionally, the flow rate may change from an oscillating flow at lower motor speeds to a non-oscillating flow at higher speeds. This affects the fluid inertia values with a factor of up to one third~\cite{doebelin1998}. This expected change in inertial flow would adversely affect the simulation at larger speeds. The discrepancies seen in Figs.~\ref{3r_150_p1}, \ref{3r_150_p2}, \ref{2r_150_p1}, and~\ref{2r_150_p2} indicate this adverse effect with larger peak values at larger flow rates. The pressure response of the three-roller configuration has average Pearson correlation coefficients ($\rho_{_\text{X,Y}}$) of 0.83 and 0.74 for the inlet and outlet respectively over all tested motor speeds. The two-roller configuration was found to have average Pearson correlation coefficients of 0.80 and 0.79 for the inlet and outlet respectively over all tested motor speeds.

\section{Conclusion}
The strong correlation of the non-linear flow rate to the inlet pressures over varying motor speeds is indicative that the slight non-linearity of the volume displacement, and hence the flow rate, are caused by water hammering effects within the hydraulic circuit. This effect can increase pressure within the tube segment between two successive roller before total occlusion occurs. This can cause the tube to swell, or possibly shrink with lower pressures, due to the tube compliance, therefore altering the amount of fluid transported and thus flow rates. 

The higher accuracy of the three-roller configuration at lower motor speeds is hypothesised to be due to the tube conformity towards the backplate. The three-roller should, in theory, hold the process tube in place better than the two-roller configuration. If the process tube does not conform toward the wall of the backplate, the tube bends away, effectively reducing the total amount of fluid stored in the tube. At larger motor speeds, the momentum of the tube would be insufficient to recoil away from the backplate before the following roller displaces the tube once more. 

The simulated pressure response reflects the experimental results rather well for motor speeds larger than 50~r/min. The proposed model is less accurate than the roller-centric model proposed in~\cite{validated}, however, allows for modelling varying numbers of rollers by altering the frequency of the controlled sources. The larger inaccuracy is attributed to simplification of the roller-centric model. The simplification of the model is partially achieved by not taking into account the tube segment between two successive rollers. The flow-centric model can therefore be improved upon by taking the tube segment between two successive rollers into account. This should, in theory, also allow for the simulation to reflect the non-linear flow rate characteristics found on the experimental results. 

Future work to investigate methods of approximating the roller volume displacement is warranted. This is in order to model and and simulate the peristaltic pump's flow rates and pressure pulsations without requiring physical measurements.

\section*{Acknowledgements}
This paper does not have any conflicts of interest and was funded by the Manufacturing, Engineering and Related services Skills, Education, and Training Authorities (merSETA) of South Africa.

\bibliography{mybibfile}

\begin{thebibliography}{10}
\expandafter\ifx\csname url\endcsname\relax
  \def\url#1{\texttt{#1}}\fi
\expandafter\ifx\csname urlprefix\endcsname\relax\def\urlprefix{URL }\fi
\expandafter\ifx\csname href\endcsname\relax
  \def\href#1#2{#2} \def\path#1{#1}\fi

\bibitem{Passaroni2015}
A.~Passaroni, M.~Silva, W.~Yoshida, Cardiopulmonary bypass: Development of john
  gibbon's heart-lung machine, Revista Brasileira de Cirurgia Cardiovascular 30
  (March 2015).
\newblock \href {https://doi.org/10.5935/1678-9741.20150021}
  {\path{doi:10.5935/1678-9741.20150021}}.

\bibitem{Mejak2000}
B.~L. Mejak, A.~Stammers, E.~Rauch, S.~Vang, T.~Viessman, A retrospective study
  on perfusion incidents and safety devices, Perfusion 15~(1) (2000) 51--61,
  pMID: 10676868.
\newblock \href {https://doi.org/10.1177/026765910001500108}
  {\path{doi:10.1177/026765910001500108}}.

\bibitem{peristaltic_pumps_review}
J.~{Klespitz}, L.~{Kovács}, Peristaltic pumps — a review on working and
  control possibilities, in: 2014 IEEE 12th International Symposium on Applied
  Machine Intelligence and Informatics (SAMI), 2014, pp. 191--194.
\newblock \href {https://doi.org/10.1109/SAMI.2014.6822404}
  {\path{doi:10.1109/SAMI.2014.6822404}}.

\bibitem{shepherd2011multigait}
R.~F. Shepherd, F.~Ilievski, W.~Choi, S.~A. Morin, A.~A. Stokes, A.~D. Mazzeo,
  X.~Chen, M.~Wang, G.~M. Whitesides, Multigait soft robot, Proceedings of the
  National Academy of Sciences 108~(51) (2011) 20400--20403.
\newblock \href {https://doi.org/10.1073/pnas.1116564108}
  {\path{doi:10.1073/pnas.1116564108}}.

\bibitem{rus2015}
D.~Rus, M.~Tolley, Design, fabrication and control of soft robots, Nature 521
  (2015) 467--75.
\newblock \href {https://doi.org/10.1038/nature14543}
  {\path{doi:10.1038/nature14543}}.

\bibitem{alfayad2011high}
S.~Alfayad, F.~B. Ouezdou, F.~Namoun, G.~Gheng, High performance integrated
  electro-hydraulic actuator for robotics – part i: Principle, prototype
  design and first experiments, Sensors and Actuators A: Physical 169~(1)
  (2011) 115 -- 123.
\newblock \href {https://doi.org/https://doi.org/10.1016/j.sna.2010.10.026}
  {\path{doi:https://doi.org/10.1016/j.sna.2010.10.026}}.

\bibitem{alfayad2011high2}
S.~Alfayad, F.~B. Ouezdou, F.~Namoun, G.~Gheng, High performance integrated
  electro-hydraulic actuator for robotics. part ii: Theoretical modelling,
  simulation, control \& comparison with real measurements, Sensors and
  Actuators A: Physical 169~(1) (2011) 124 -- 132.
\newblock \href {https://doi.org/https://doi.org/10.1016/j.sna.2011.03.019}
  {\path{doi:https://doi.org/10.1016/j.sna.2011.03.019}}.

\bibitem{sideris2020pumps}
E.~Sideris, H.~{de Lange}, Pumps operated by solid-state electromechanical
  smart material actuators - a review, Sensors and Actuators A: Physical 307
  (2020) 111915.
\newblock \href {https://doi.org/https://doi.org/10.1016/j.sna.2020.111915}
  {\path{doi:https://doi.org/10.1016/j.sna.2020.111915}}.

\bibitem{weinberg1971experimental}
S.~L. Weinberg, E.~C. Eckstein, A.~H. Shapiro, An experimental study of
  peristaltic pumping, Journal of Fluid Mechanics 49~(3) (1971) 461--479.
\newblock \href {https://doi.org/10.1017/S0022112071002209}
  {\path{doi:10.1017/S0022112071002209}}.

\bibitem{latham1966fluid}
T.~W. Latham, \href{http://hdl.handle.net/1721.1/17282}{Fluid motions in a
  peristaltic pump}, Ph.D. thesis, Massachusetts Institute of Technology,
  Cambridge, MA (1966).
\newline\urlprefix\url{http://hdl.handle.net/1721.1/17282}

\bibitem{SALAHUDDIN20208337}
T.~Salahuddin, M.~Habib, M.~Arshad, M.~Abdel-Sattar, Y.~Elmasry, Peristaltic
  transport of {$\gamma$Al$_2$O$_3$/H$_2$O and
  $\gamma$Al$_2$O$_3$/C$_2$H$_6$O$_2$} in an asymmetric channel, Journal of
  Materials Research and Technology 9~(4) (2020) 8337--8349.
\newblock \href {https://doi.org/https://doi.org/10.1016/j.jmrt.2020.05.012}
  {\path{doi:https://doi.org/10.1016/j.jmrt.2020.05.012}}.

\bibitem{validated}
F.~Moscato, F.~M. Colacino, M.~Arabia, G.~A. Danieli, Pressure pulsation in
  roller pumps: A validated lumped parameter model, Medical Engineering \&
  Physics 30 (2008) 1149--1158.
\newblock \href {https://doi.org/10.1016/j.medengphy.2008.02.007}
  {\path{doi:10.1016/j.medengphy.2008.02.007}}.

\bibitem{karassik2000pump}
I.~Karassik, J.~Messina, P.~Cooper, C.~Heald,
  \href{https://books.google.co.za/books?id=yU5TyJrOMF8C}{Pump Handbook},
  Harvard Business Review Book Series, Mcgraw-hill, 2000.
\newline\urlprefix\url{https://books.google.co.za/books?id=yU5TyJrOMF8C}

\bibitem{McIntyre2020}
M.~P. McIntyre, Modelling and characterisation of a {3D} printed peristaltic
  pump, Master's dissertation, North-West University of South Africa,
  Potchefstroom, North-West (2020).

\bibitem{karnopp2012system}
D.~C. Karnopp, D.~L. Margolis, R.~C. Rosenberg, System Dynamics: Modeling,
  Simulation, and Control of Mechatronic Systems, John Wiley \& Sons, 2006.

\bibitem{Menon2015}
E.~Menon, Piping Calculations Manual, 1st Edition, McGraw-Hill, 2015.

\bibitem{schober2018correlation}
P.~Schober, C.~Boer, L.~A. Schwarte, Correlation coefficients: appropriate use
  and interpretation, Anesthesia \& Analgesia 126~(5) (2018) 1763--1768.
\newblock \href {https://doi.org/https://doi.org/10.1213/ANE.0000000000002864}
  {\path{doi:https://doi.org/10.1213/ANE.0000000000002864}}.

\bibitem{doebelin1998}
E.~Doebelin, System Dynamics: modelling, analysis, simulation, design, Marcel
  Dekker, Inc., 1998.

\end{thebibliography}

\end{document}